\documentclass[%
 11pt,
superscriptaddress,
 amsmath,amssymb,
 aps,
]{revtex4-2}

\usepackage{mathtools}

\usepackage{graphicx}% Include figure files
\usepackage{dcolumn}% Align table columns on decimal point
\usepackage{bm}% bold math
\usepackage{color}
\newcommand{\red}[1]{{\color{black}#1}}

\def\ii{\mathbf{i}}
\def\j{\mathbf{j}}
\def\e{\varepsilon}

\begin{document}

\title{Stability of nanoparticle laden aerosol liquid droplets}

\author{A.J.~Archer}
\email{a.j.archer@lboro.ac.uk}
 \affiliation{Department of Mathematical Sciences and Interdisciplinary Centre for Mathematical Modelling, Loughborough University, Loughborough LE11 3TU, UK}%Lines break automatically or can be forced with \\
\author{B.D.~Goddard}%
 \email{b.goddard@ed.ac.uk}
\affiliation{School of Mathematics and the Maxwell Institute for Mathematical Sciences, University of
Edinburgh, Edinburgh EH9 3FD, UK}%

\author{R.~Roth}
\email{roland.roth@uni-tuebingen.de}
\affiliation{Institute for Theoretical Physics, University of T\"ubingen, 72076 T\"ubingen, Germany}%

\date{\today}% It is always \today, today,

\begin{abstract}
We develop a model for the thermodynamics and evaporation dynamics of aerosol droplets of a liquid such as water, surrounded by the gas. When the temperature and the chemical potential (or equivalently the humidity) are such that the vapour phase is the thermodynamic equilibrium state, then of course droplets of the pure liquid evaporate over a relatively short time. However, if the droplets also contain nanoparticles or any other non-volatile solute, then the droplets can become thermodynamically stable. We show that the equilibrium droplet size depends strongly on the amount and solubility of the nanoparticles within, i.e.\ on the nature of the particle interactions with the liquid, and of course also on the vapour temperature and chemical potential. We develop a simple thermodynamic model for such droplets and compare predictions with results from a lattice density functional theory that takes as input the same particle interaction properties, finding very good agreement. We also use dynamical density functional theory to study the evaporation/condensation dynamics of liquid from/to droplets as they equilibrate with the vapour, thereby demonstrating droplet stability.
\end{abstract}

\maketitle

\section{Introduction}
An aerosol droplet is a small liquid drop in the colloidal size range of tens of nanometers up to a few micrometers in diameter, suspended in a gas like air. The lifetime of such an aerosol droplet is determined by a competition between gravity and evaporation \cite{wells1934,xie2007}. Larger droplets sediment from a height of the order of 2 meters in less than a few seconds, while smaller droplets evaporate rapidly, as long as the temperature and pressure conditions are such that the vapour of the volatile liquid is the thermodynamic equilibrium phase. In the case of water, such small droplets evaporate completely in the order of seconds. Wells, who was interested in the spreading of diseases through aerosols \cite{wells1934} showed that the average sedimentation time scales with $1/R_0^2$, where $R_0$ is the initial droplet radius, while the evaporation time scales with $R_0^2$. Thus, a droplet of pure water always either sediments or evaporates sufficiently quickly, preventing it from travelling a significant horizontal distance from the source. In contrast to droplets described by these simple estimates, aerosols produced by people when they breathe, talk, cough or sneeze can stay air-borne and thus potentially dangerous for much longer times, of the order of hours instead of seconds.

This is because, if the droplets contain nanoparticles, e.g.\ pathogenic germs, suspended within the liquid or some other non-volatile solute, then a sufficient quantity of these can stabilize the droplets for hours or even indefinitely \cite{netz2020mechanisms, seyfert2022stability}. In other words, once such a droplet is formed, there may be some fast initial evaporation of the solvent liquid, but in the long time limit, a finite-size droplet remains that can stay air-borne. The droplet is stabilised by the sufficiently large concentration of nanoparticles within. In the equilibrium state, the rate at which liquid evaporates from this droplet is balanced by condensation of the vapour phase onto the droplet. Clearly, the temperature and relative humidity of the vapour atmosphere have to be taken into account \cite{netz2020mechanisms, seyfert2022stability}.

Thus, studying such droplets allows  to address questions such as how long does a droplet of saliva survive in the air before it evaporates? How does this depend on the amount of solute within? These questions are particularly relevant if any of that solute is an infectious virus, such as COVID or influenza. A related additional question of particular interest here is: what influence do the interactions between the non-volatile material within and the surrounding liquid have on the droplet size? Saliva droplets contain a significant amount of material other than water, including various mucus components, proteins, salts and sometimes virus particles \cite{poon2020soft, seyfert2022stability, vejerano2018physico}. That this non-volatile material can stabilize the droplets goes some way to explaining why it was observed that influenza viruses remain stable and infectious in aerosols across a wide range of relative humidities~\cite{kormuth2018influenza}. Note that the air humidity is directly related to the vapour pressure or, equivalently, is determined by the chemical potential of the vapour. Because these quantities are so closely connected, we refer to them here almost interchangeably.

Aerosol droplets are not just relevant to the spread of disease: They are ubiquitous in the global environment, playing a crucial role in atmospheric science and meteorology. For example, aerosol droplets play a role in determining the how long clouds persist in the sky, before the water they contain returns to the ground as rain \cite{stevens2009untangling}. They are also often the host locations for chemical reactions between air-borne species \cite{wei2018aerosol} and the lifetime and stability of aerosol droplets must be considered when addressing the question of how long harmful chemicals and other pollutants remain in the atmosphere \cite{von2015chemistry}.

In what follows, we refer to all of the non-volatile material that may be found in an aerosol liquid droplet collectively as `nanoparticles'. Droplets of pure water evaporate in the air, but droplets containing sufficient numbers of nanoparticles are stabilised by the particles within.
The recent study by Netz \cite{netz2020mechanisms} addressed this and many of the related issues to do with the evaporation of saliva droplets in air. When considering the thermodynamic stability of saliva droplets, Netz used a simple thermodynamics for mixtures, based on assuming ideal-mixing and ideal-volume additivity. Here, we go beyond this to develop a model to determine the equilibrium size of such droplets and how the droplet size depends on the humidity and temperature of the surrounding gas and also on the nature of the interactions of the solute particles with the solvent liquid. We show that nanoparticles that have a higher solubility (i.e.\ have a greater preference for being dispersed in the liquid) lead to larger droplets. We develop a simple capillarity-approximation based thermodynamic model that we validate by comparing with the results from a lattice density functional theory (DFT), which is a theory for the microscopic density profile of the liquid and nanoparticles within the droplet \cite{evans1979nature, hansen2013theory}. We also study the evaporation/condensation dynamics of the liquid from/onto non-equilibrium droplets using dynamical density functional theory (DDFT) \cite{hansen2013theory, marconi1999dynamic,archer2006dynamical,te2020classical}, thereby demonstrating the stability of the droplets and elucidating properties of the formation dynamics. \red{By using DDFT, we assume that the motion of all particles in the system is diffusive, stemming from our expectation that the droplet dynamics is dominated by diffusive exchange of solvent molecules from the liquid to the vapour phase and also that all motion within the droplet can be treated as diffusive.} In our work here, we treat the surrounding gas as being solely made up of the vapour phase and do not treat explicitly the inert gas molecules that are present, e.g.\ in the atmosphere. However, since these largely play the role of passive spectators, the results presented here do also apply to the case of aerosol water droplets in air.

The lattice DFT that we use here and variants of it have previously been applied successfully to the study of various properties of \red{liquids condensing and adsorbed in pores and porous media \cite{kierlik2001capillary, woo2001mean, schneider2014filling} and to} liquid droplets on surface \cite{hughes2014introduction, hughes2015liquid, chalmers2017dynamical, areshi2019kinetic}. The recent study in Ref.~\cite{perez2021changing} is particularly noteworthy because comparisons with experimental measurements on water droplets containing nanoparticles drying on various surfaces was made, demonstrating good agreement between the theory and the experiments. In view of this, we are confident that the lattice DFT and DDFT used here also provides a good description of the thermodynamics and dynamics of aerosol droplets. Additionally, in a future study, we will support our lattice DFT results by additional off-lattice DFT calculations.

This paper proceeds a follows: In Sec.~\ref{sec:model} we develop a generalised lattice-gas model for nanoparticle laden droplets and present the lattice DFT that we use to determine the density profiles of the liquid and nanoparticles within the droplets. This theory also yields all the relevant thermodynamic quantities pertaining to the droplets. Bulk thermodynamic quantities are also briefly discussed in this section. In Sec.~\ref{sec:equilib_DFT} we present results for the liquid and nanoparticle density profiles in droplets calculated using our DFT model. In Sec.~\ref{sec:cap_model} we then present our capillarity-approximation based thermodynamic model, comparing with DFT results for the total amount of liquid in the droplets, as quantities like the number of nanoparticles, humidity and interaction strength between the nanoparticles and the liquid are varied. In Sec.~\ref{sec:dynamics} we present our DDFT model and results for the formation dynamics of droplets. Then, in Sec.~\ref{sec:DDFT_v_Picard} we compare results for droplet dynamics obtained from DDFT with the dynamics from Picard iteration, which is a fictitious dynamics used to solve the equations of DFT. This section will be of interest to DFT practitioners, but may be skipped by those from a more general audience. Finally, in Sec.~\ref{sec:conc} we make a few concluding remarks.

\section{Model for nanoparticle laden droplets}
\label{sec:model}

\begin{figure}[t] %  figure placement: here, top, bottom, or page
   \centering
   \includegraphics[width=2.6in]{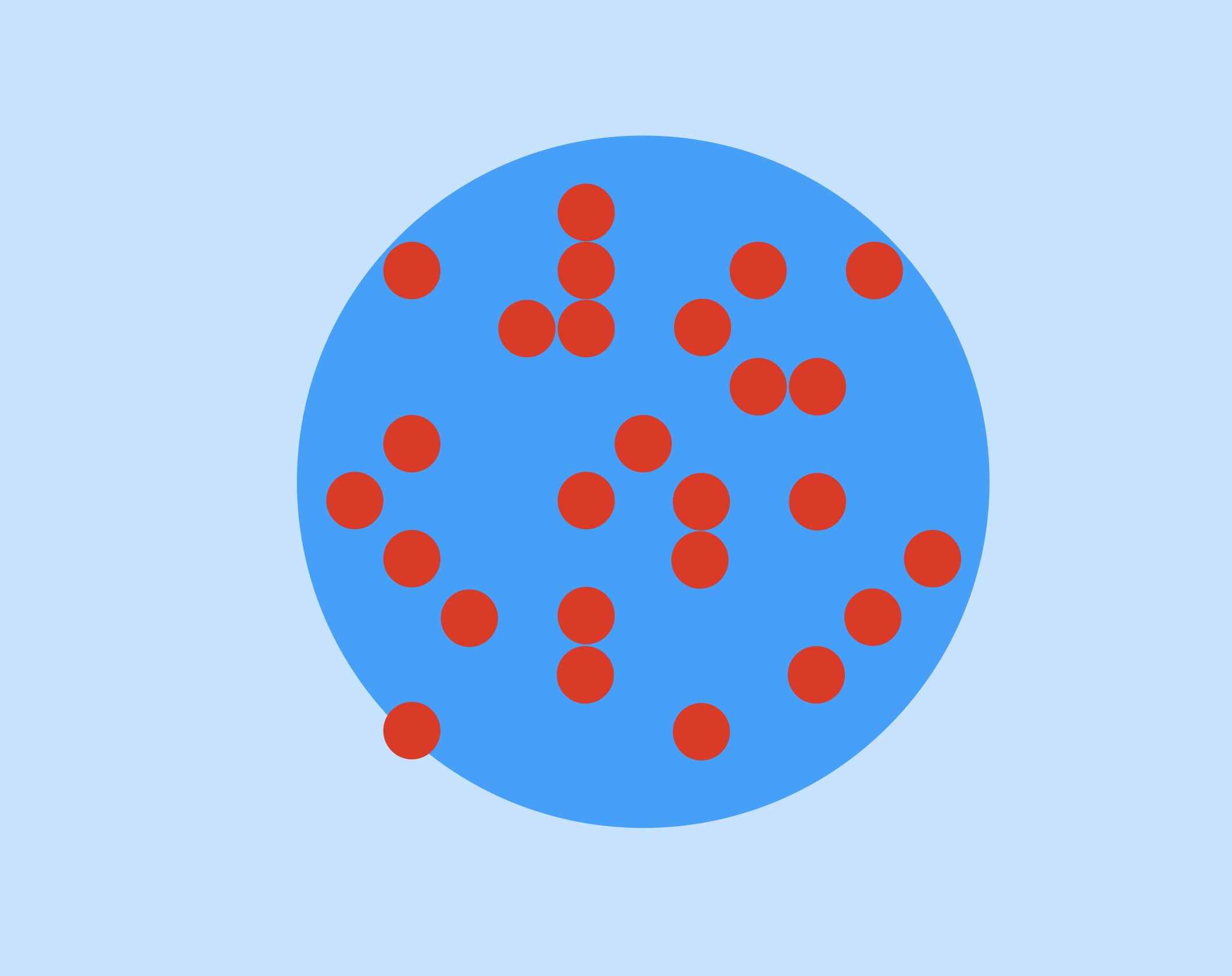} 
   \includegraphics[width=2.5in]{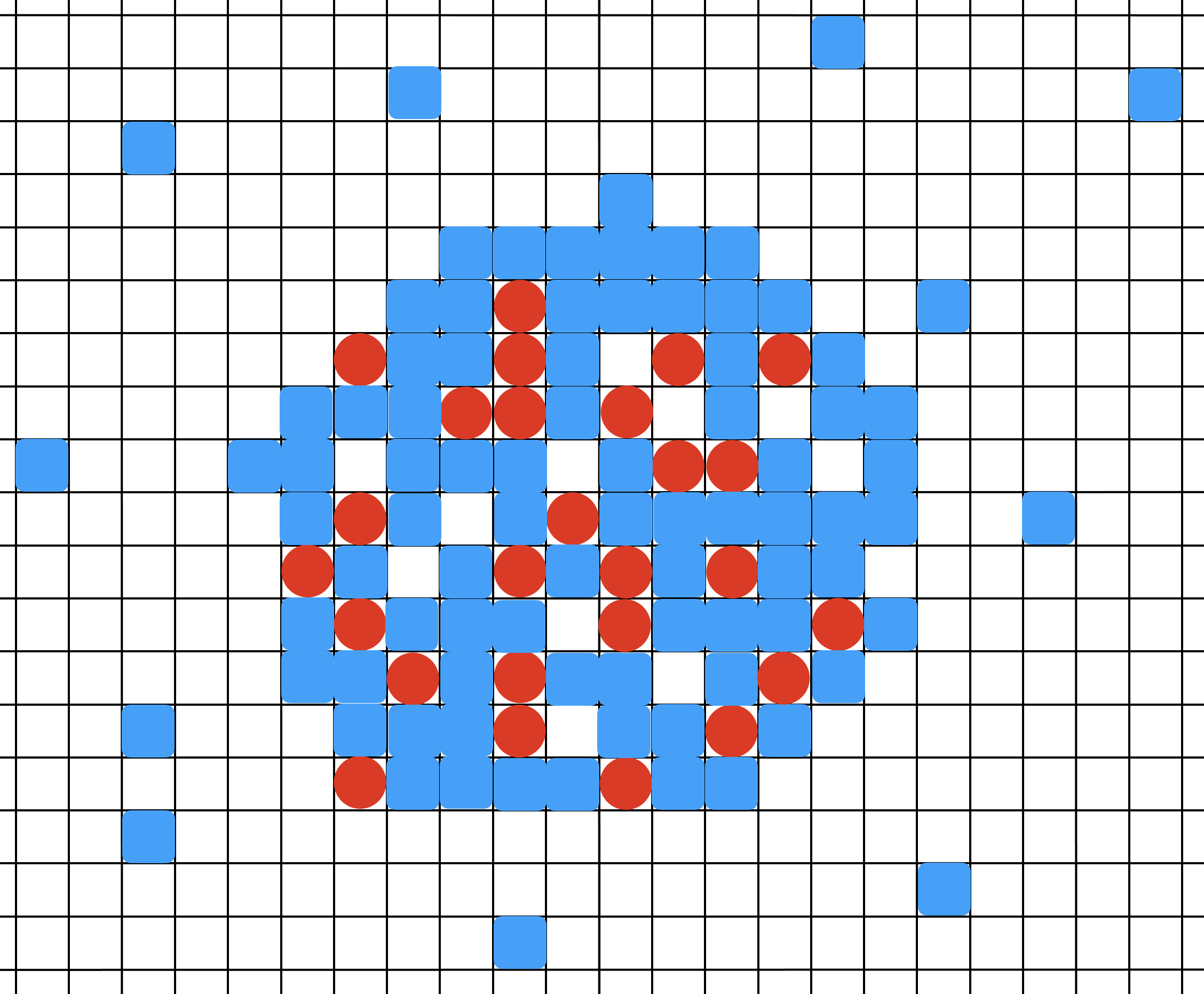} 
   \caption{On the left is an illustration of the system of interest, namely an aerosol liquid droplet containing nanoparticles that is surrounded by gas (i.e.\ the vapour phase). The red circles represent the suspended nanoparticles. To treat this, we coarse-grain the system onto a \red{square} lattice, as illustrated on the right. We choose the size of each lattice site to correspond roughly to the diameter of the nanoparticles. Lattice sites are described as either occupied with a nanoparticle (red circle) or occupied by liquid (blue square) or empty (white). \red{We also define effective interaction potentials between pairs of lattice sites that represent coarse-grained analogues of the inter-particle interaction potentials in the original system on the left.} Thus, the liquid within the drop has the majority of lattice sites full of either liquid or nanoparticles, while the vapour outside has most (but not all) sites being empty.}
   \label{fig:diagram}
\end{figure}

In Fig.~\ref{fig:diagram} we illustrate a cross-section through an aerosol liquid droplet containing nanoparticles. To model this, we discretise the system onto a lattice, as also illustrated in Fig.~\ref{fig:diagram}. We choose the lattice spacing to be the diameter of the nanoparticles,~$\sigma$. Thus, lattice sites that contain a nanoparticle have just one of them within, while lattice sites that are `full' of the liquid, have a large number of the liquid molecules within. \red{As an example of a concrete physical system, consisting of the case where the solvent is water and for $\sigma\approx100$nm (roughly the diameter of a COVID viron), we would find $\approx 3.3\times10^7$ molecules in each lattice site `full' of the water.} Following Ref.~\cite{chalmers2017dynamical}, this discretization onto a lattice allows us to map the system onto a two-species lattice-gas (generalised Ising model) and thereby to investigate the thermodynamics of the system. The resulting model can also be thought of as a discretised partial differential equation for the density distributions of the liquid and nanoparticles~\cite{robbins2011modelling}.

We denote the location of each lattice site by the index $\ii$. In three dimensions, we have $\ii =(i,j,k)$, where $i$, $j$ and $k$ are integers. To simplify our calculations below, we assume instead that our system is two-dimensional (2D) and therefore we have $\ii =(i,j)$. We introduce two occupation numbers for each lattice site, $p_\ii^n$ and $p_\ii^l$, for the nanoparticles and for the liquid, respectively. If a lattice site is empty, then both $p_\ii^n=p_\ii^l=0$. If lattice site $\ii$ is occupied by a nanoparticle, then $p_\ii^n=1$, while $p_\ii^l=0$. If instead lattice site $\ii$ is full of liquid, then $p_\ii^n=0$ and $p_\ii^l=1$. We assume that it is impossible for both occupation numbers to simultaneously equal one. The potential energy of the system can then be approximated as
\begin{align}
\label{eq:Hamiltonian}
E = - \sum_{\ii,\j}\left(\frac{1}{2} \e_{\ii\j}^{ll} p_{\ii}^l p_{\j}^l + \e_{\ii\j}^{nl} p_\ii^l p_\j^n + \frac{1}{2} 
\e_{\ii\j}^{nn} p_\ii^n p_\j^n\right) 
%\nonumber
%\\
%- \mu_l \sum_\ii p_\ii^l - \mu_n \sum_\ii p_\ii^n
+ \sum_\ii \Phi_\ii^l \,p_\ii^l + \sum_\ii \Phi_\ii^n p_\ii^n,
\end{align}
where the first three terms arise from the interactions between pairs of occupied lattice sites, while the last two terms are the contributions from any external potentials $\Phi_\ii^l$ and $\Phi_\ii^n$ on the liquid and nanoparticles, respectively. Here, we assume $\Phi_\ii^l=\Phi_\ii^n=0$. Note that $\sum_\ii=\sum_{i=1}^{M_x}\sum_{j=1}^{M_y}$, i.e.\ this denotes a sum over all lattice sites in the system, where $M_x$ is the number of lattice sites along the Cartesian $x$-direction indexed by $i$ and $M_y$ is the number along the $y$-direction, indexed by $j$. Similarly, $\sum_{\ii,\j}$ denotes a sum over all pairs of lattice sites. The pair interaction terms involved the discretised pair-potentials $\e_{\ii\j}^{ll}$, $\e_{\ii\j}^{nl}$ and $\e_{\ii\j}^{nn}$, which we assume are given by the matrices $\e_{\ii\j}^{ll}=\e^{ll}c_{\ii\j}$, $\e_{\ii\j}^{nl}=\e^{nl}c_{\ii\j}$ and $\e_{\ii\j}^{nn}=\e^{nn}c_{\ii\j}$, where the parameters $\e^{ll}$, $\e^{nl}$ and $\e^{nn}$ determine the overall strength of the pair-interactions and the matrix
\begin{equation}
\label{eq:c_ij}
c_{\ii\j}  = 
  \begin{cases} 
   1 & \text{if }\j\in {NN \ii}, \\
    \frac{1}{2}    & \text{if }\j\in {NNN \ii}, \\
        0   & \text{otherwise},
         \end{cases}
\end{equation}
where ${NN \ii}$ denotes the lattice sites that are nearest neighbours to site $\ii$, while ${NNN \ii}$ indicates the sites that are the next nearest neighbours. These three short-ranged (truncated) potentials are chosen to have the above form for the reasons discussed in \cite{robbins2011modelling, chalmers2017dynamical}, namely that this choice greatly reduces the influence of the approximation of having discretised the system onto a lattice. If other values for the entries of the matrix $c_{\ii\j}$ were used, we would obtain e.g.\ non-circular droplets and other similar influences from the underlying grid. There is a corresponding choice one can make when applying the model in 3D \cite{chalmers2017modelling, chalmers2017dynamical, areshi2019kinetic}. Note that with the sign convention used in Eq.~\eqref{eq:Hamiltonian} for the pair interaction terms, positive values of $\e^{ll}$, $\e^{nl}$ and $\e^{nn}$ correspond to attractive interactions between neighbouring particles, while negative values (not considered here) correspond to repulsive interactions. The parameter $\e^{nl}$ determines the overall strength of attraction between a nanoparticle and any liquid surrounding it. Thus, varying this parameter most directly determines the free energy to insert a nanoparticle into the liquid \cite{chipot2007free, roth2006morphometric, coe2023understanding}, which is the quantity that determines the solubility of the nanoparticles in the liquid.

\subsection{Lattice DFT and thermodynamics}
\label{sec:DFT_thermo}

Having defined the Hamiltonian~\eqref{eq:Hamiltonian}, one could proceed e.g.\ by performing Monte-Carlo computer simulations, as was done in Ref.~\cite{chalmers2017modelling}. However, here we prefer to follow e.g.\ Refs.~\cite{hughes2014introduction, hughes2015liquid, chalmers2017dynamical} to develop a theory for the ensemble-averaged densities
\begin{align}
\rho_{\ii}^l = \langle p_\ii^l\rangle \,\,\,\,\text{and} \,\,\,\,\rho_\ii^n = \langle p_\ii^n\rangle.
\end{align}
Thus, we apply an extension of DFT \cite{hansen2013theory, evans1979nature} to lattice-systems. We use the following approximation for the Helmholtz free energy of the system \cite{woywod2003phase, chalmers2017dynamical}:
\begin{align}
F(\{\rho^l_\ii,\rho^n_\ii\})%
  &= k_B T \sum_\ii \left[
    \rho^l_\ii \ln \rho^l_\ii
    + (1 - \rho^l_\ii - \rho^n_\ii) \ln (1 - \rho^l_\ii - \rho^n_\ii)
    + \rho^n_\ii \ln \rho^n_\ii
  \right]
  \notag\\
  &\quad
    - \frac12 \sum_{\ii,\j} \e^{ll}_{\ii\j} \rho^l_\ii \rho^l_\j
    - \sum_{\ii,\j} \e^{ln}_{\ii\j} \rho^l_\ii \rho^n_\j
    - \frac12 \sum_{\ii,\j} \e^{nn}_{\ii\j} \rho^n_\ii \rho^n_\j
  %\notag\\
  %&\quad
    + \sum_\ii \left(
        \Phi^l_\ii \rho^l_\ii
      + \Phi^n_\ii \rho^n_\ii
    \right),
\label{eq:helmholtz}
\end{align}
where $k_B$ is Boltzmann's constant and $T$ is the temperature. To determine the equilibrium density profiles $\rho^l_\ii$ and $\rho^n_\ii$ corresponding to an aerosol droplet, we minimise the free energy $F$ in the semi-grand ensemble, i.e.\ with fixed total number of nanoparticles in the system
 \begin{equation}
 N_n=\sum_\ii\rho^n_\ii,
 \label{eq:Nn}
 \end{equation}
 but with the chemical potential of the liquid $\mu\equiv\mu^l$ (or equivalently the relative humidity) as an input, so that the total amount of liquid in the system
  \begin{equation}
 N_l=\sum_\ii\rho^l_\ii,
 \label{eq:N_l}
 \end{equation}
 is an output of our calculations. We calculate the equilibrium density profiles using a Picard iteration scheme similar to that described in Ref.~\cite{hughes2014introduction}. Since we are in the semi-grand ensemble, the equilibrium density profiles $\{\rho_\ii^l,\rho_\ii^n\}$ are those that minimize the semi-grand free energy
\begin{equation}
    \Omega = F - \sum_\ii \mu \rho_\ii^l,
\end{equation}
subject to the constraint that $N_n$, given by \eqref{eq:Nn}, is equal to the desired value, $\hat{N}_n$.
In other words, we solve the coupled set of equations
\begin{align}
    \frac{\partial \Omega}{\partial\rho_\ii^l} &= 0 \\
    \frac{\partial \Omega}{\partial\rho_\ii^n} &= 0 \\
    N_n &= \hat{N}_n.
\end{align}
Differentiating $\Omega$ with respect to $\rho_\ii^l$ and $\rho_i^n$ and rearranging gives the conditions
\begin{align}
    \rho_\ii^l &= (1 - \rho_\ii^l - \rho_\ii^n) 
                \exp\Big[\beta \Big( \frac{1}{2}\sum_\j (\epsilon_{\ii\j}^{ll} + \epsilon_{\j\ii}^{ln}) \rho_\ii^l + \sum_\j \epsilon_{\ii\j}^{ln}\rho_\j^n - \Phi_\ii^l + \mu \Big) \Big] \label{eq:rholIter}\\
    \rho_\ii^n &= (1 - \rho_\ii^l - \rho_\ii^n) 
                \exp\Big[\beta \Big( \frac{1}{2}\sum_\j (\epsilon_{\ii\j}^{nn} + \epsilon_{\j\ii}^{ln}) \rho_\ii^n + \sum_\j \epsilon_{\ii\j}^{ln}\rho_\j^l - \Phi_\ii^n \Big) \Big]
                \label{eq:rhonIter}\\ 
    N_n &= \hat{N}_n.\label{eq:N=N}
\end{align}
We solve this set of equations iteratively, starting from an initial guess for the profiles.  A standard Picard solver would take the current state $\rho_\ii^{m,\textrm{old}}$ and replace it with the result of evaluating the right hand sides of Eqs.~\eqref{eq:rholIter} and \eqref{eq:rhonIter} with these densities, denoted $\rho_\ii^{m,\textrm{rhs}}$. Here, and below, $m \in \{l,n\}$.
Since we are working in the semi-grand ensemble where $N_n$ is fixed, we enforce this by renormalising the density profile of the nanoparticles $\{\rho_\ii^n\}$ at each step, so that Eq.~\eqref{eq:N=N} is satisfied.
Note also that it is often necessary to mix the results from the previous step $\rho_\ii^{m,\textrm{old}}$ with the result from evaluating the right hand sides, $\rho_\ii^{m,\textrm{rhs}}$:
\begin{equation}
    \rho_\ii^{m,\textrm{new}} = \alpha \rho_\ii^{m,\textrm{rhs}}
    + (1-\alpha) \rho_\ii^{m,\textrm{old}},
\end{equation}
where $\alpha$ is typically small, e.g., $0.01 \leq \alpha \leq 0.1$.  This mixing increases the robustness of the scheme, in particular by preventing $\rho_\ii^{m,\textrm{new}}$ from lying outside of the range $(0,1)$.

Before presenting results from this lattice-DFT model, we briefly discussing a few relevant aspects of the bulk thermodynamics and phase behaviour of the system. From Eq.~\eqref{eq:helmholtz} we find that for a uniform system with $\rho_\ii^l=\rho^l=N_l/V$ and $\rho_\ii^n=\rho^n=N_n/V$, constants for all $\ii$, the Helmholtz free energy per unit volume, $f = F/V$, where $V$ is the volume of the system, is given by:
\begin{align}
\label{eq:f_bulk}
f = & k_BT[\rho^l \ln{\rho^l} + (1 - \rho^l - \rho^n ) \ln(1 - \rho^l - \rho^n) + \rho^n \ln{\rho^n}]
\nonumber
\\ 
& - 3 \e^{ll}(\rho^l)^2 - 6 \e^{nl} \rho^l \rho^n - 3 \e^{nn} (\rho^n)^2.
\end{align}
From this, we can obtain the following relation between the chemical potential of the liquid and the particle densities,
\begin{equation}
\mu = \frac{\partial f}{\partial \rho^l}.
\label{eq:mu}
\end{equation}
The pressure (or equivalently minus the semi-grand potential density) is obtained from the relation
\begin{equation}
p = -f+\mu\rho^l \label{eq:p}.
\end{equation}
For the case where no nanoparticles are present, i.e.\ where $\rho^n\to0$, the system exhibits vapour-liquid phase separation. The critical point occurs for $\beta\e^{ll}=2/3$, where $\beta=(k_BT)^{-1}$; i.e.\ the critical point is at $(\rho^l,T)=(\rho^l_c,T_c)=(\frac{1}{2},\frac{3\e^{ll}}{2k_B})$. Bulk liquid-vapour phase coexistence occurs for $\mu=-3\e^{ll}$. Adding nanoparticles to the system can in general completely change the bulk phase behaviour -- see e.g.\ Ref.~\cite{areshi2020mathematical}. However, for the interaction parameter values considered here, generally the effect of the nanoparticles is to just shift somewhat the coexisting density and chemical potential values, but the overall qualitative phase behaviour remains the same.

\section{Equilibrium droplet density profiles}
\label{sec:equilib_DFT}

\begin{figure}[ht!] %  figure placement: here, top, bottom, or page
   \centering
    \includegraphics[width=0.7\textwidth]{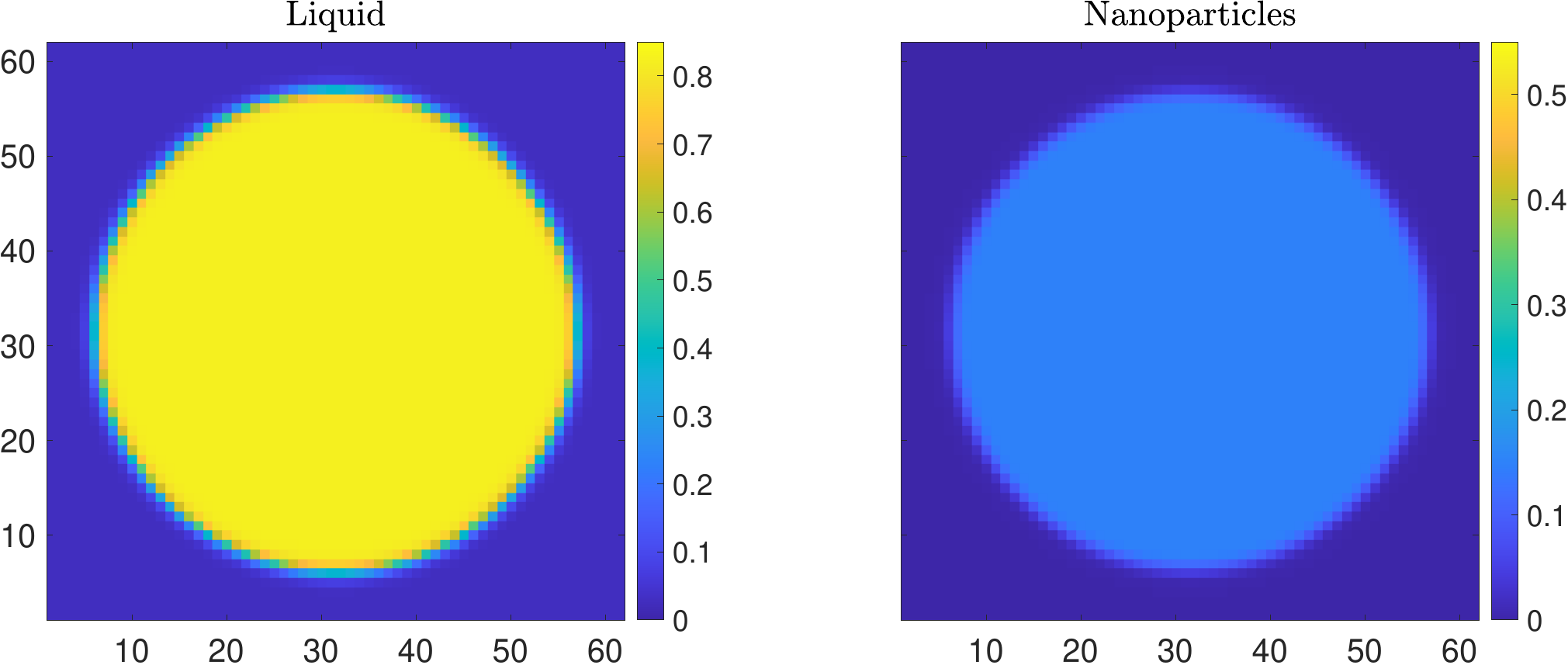}\\
    \includegraphics[width=0.7\textwidth]{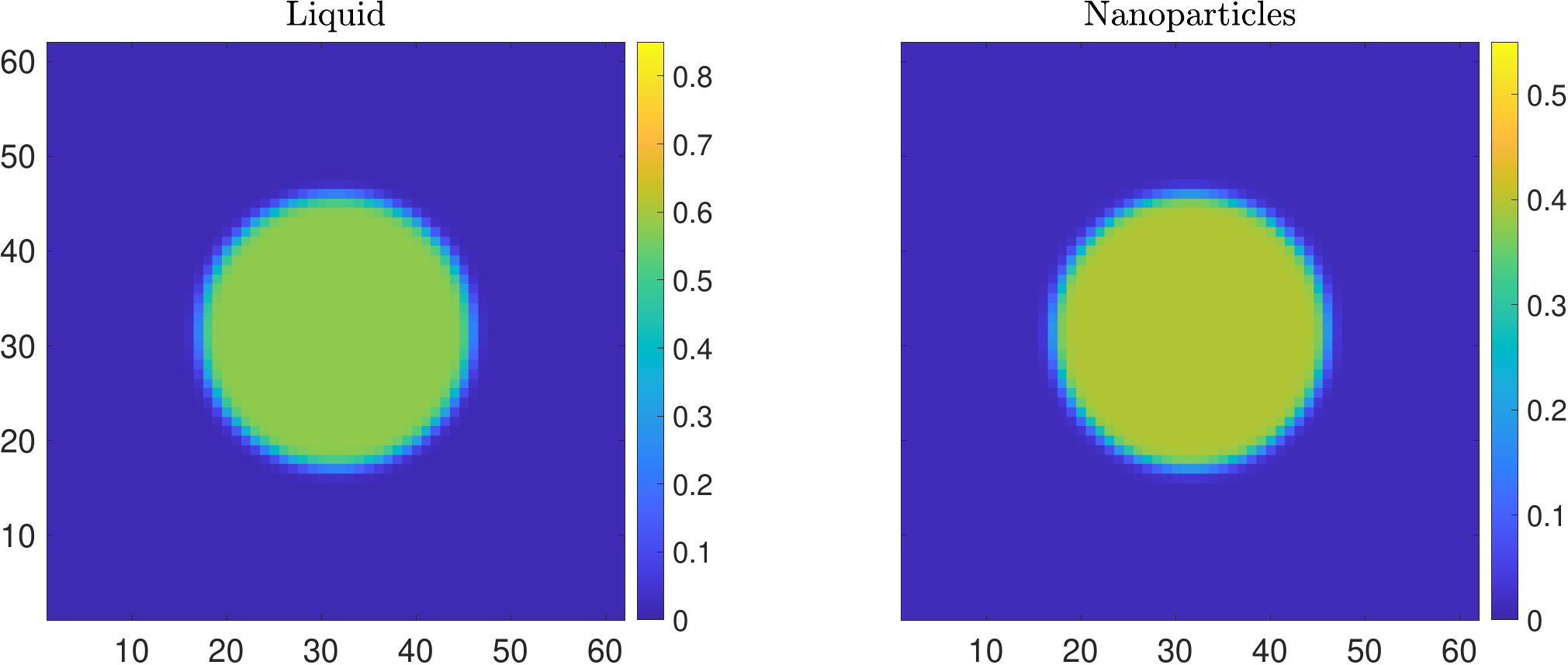}\\
    \includegraphics[width=0.7\textwidth]{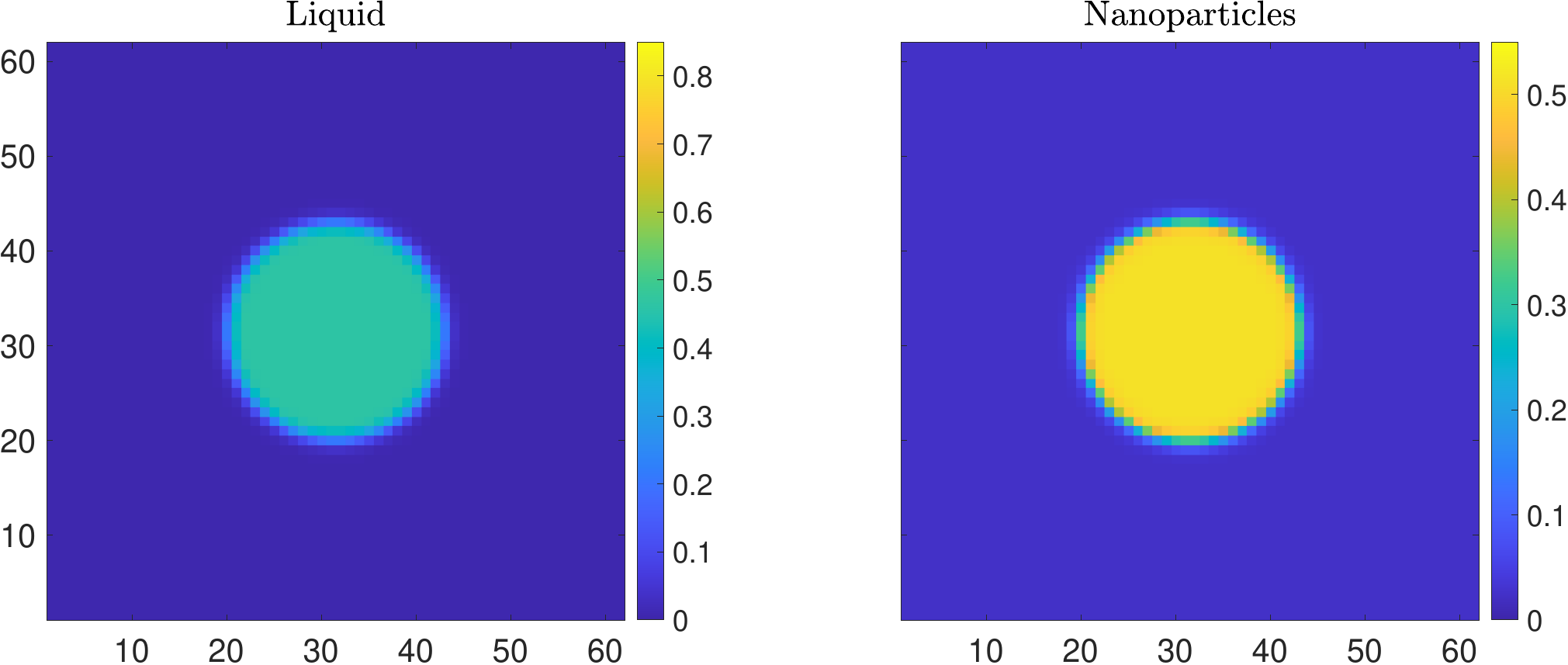}
   \caption{Top row: the density profiles (liquid on the left, nanoparticles on the right), for an equilibrium nanoparticle laden droplet. The system interaction parameters are $\beta\e^{ll}=1.2$, $\beta\e^{nn}=0.9$, $\beta\e^{nl}=1.5$ and liquid chemical potential $\beta\mu=-3.8$ (corresponding to moist air with $H_r=80\%$), while the total number of nanoparticles in the system is constrained to be $N_n=300$. Middle row: the corresponding case for $\beta\mu=-4.5$ (comfortable moisture level, with $H_r=38\%$), and all other parameters the same. Bottom row: the corresponding case for $\beta\mu=-5$ (dry air, with $H_r=23\%$), again with all other parameters the same. Note that the droplet decreases in size with decreasing $\mu$, i.e.\ with decreasing air humidity. Moreover, the liquid density within the droplet is also less.}
   \label{fig:profiles}
\end{figure}

As mentioned above, to simplify our DFT calculations, we treat the system as being 2D, but our results could fairly easily be repeated in 3D, at the expense of more computational cost. We consider the system with $\beta\e^{ll}=1.2$, where the densities of the coexisting (pure) liquid and gas are $\rho^l=0.034$ and $\rho^l=0.966$. For the interaction strengths with the nanoparticles, we initially set $\beta\e^{nn}=0.9$ and $\beta\e^{nl}=1.5$. This choice has $\e^{nn}<\e^{ll}<\e^{nl}$, which ensures that the nanoparticles prefer to stay well-dispersed within the liquid and do not aggregate. If $\e^{nn}$ were larger and/or if $\e^{nl}$ were smaller, this could potentially lead to the nanoparticles and the liquid demixing  \cite{areshi2020mathematical}. In Sec.~\ref{sec:cap_model} below, we present results for different values of $\e^{nl}$, which is the parameter that most directly determines the solubility of the nanoparticles in the liquid.

In Fig.~\ref{fig:profiles} we present both the liquid and nanoparticle density profiles for the case when the total number of nanoparticles in the system is set to be $N_n=300$, in a square box of size $55\sigma\times55\sigma$. In the top row are displayed results for the case when $\beta\mu=-3.8$, in the middle row for $\beta\mu=-4.5$ and in the bottom row for $\beta\mu=-5$. For the temperature considered here (i.e.\ for the value of $\beta\e^{ll}=1.2$, used here), the chemical potential at gas-liquid coexistence is $\beta\mu_{coex}=-3.6$ and so these three chemical potential values correspond to relative humidity values of $H_r=80\%$, $H_r=38\%$ and $H_r=23\%$, respectively. If this were for the case of water in the air, then these $H_r$ values correspond roughly speaking a moist environment, a typical comfortable day-to-day level and to a fairly dry level, where one may need extra moisturising cream on skin. Note that the relative humidity is defined as $H_r(\mu)=100\times p(\mu)/p(\mu_{coex})$, where the corresponding pressures are calculated via Eqs.~\eqref{eq:f_bulk}--\eqref{eq:p}. We see that when the chemical potential is closer to the value at phase-coexistence (more humid air), the equilibrium droplet size is much larger than in the dry air, which of course is to be expected.

\begin{figure}[ht!] %  figure placement: here, top, bottom, or page
   \centering
    \includegraphics[width=0.7\textwidth]{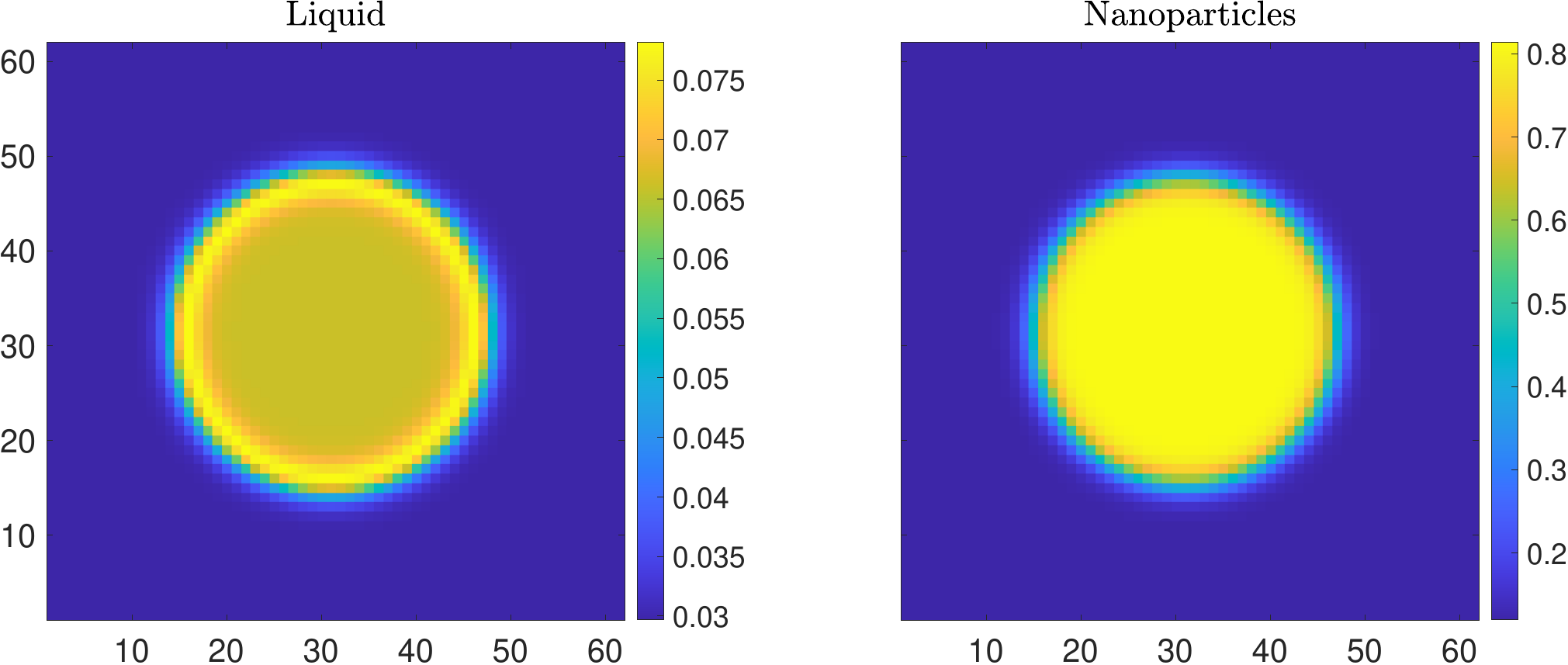}
   \caption{Density profiles (liquid on the left, nanoparticles on the right), for the case when $\beta\e^{ll}=1.2$, $\beta\e^{nn}=0.9$, $\beta\e^{nl}=0.6$ and liquid chemical potential $\beta\mu=-4$ (corresponding to $H_r=65\%$), while the total number of nanoparticles in the system is constrained to be $N_n=1000$. The smaller value of $\e^{nl}$ used here (compared to that used in Fig.~\ref{fig:profiles}), results in a high nanoparticle density within the drop and interestingly the liquid density profile is highest on the edge of the droplet.}
   \label{fig:profiles2}
\end{figure}

In Fig.~\ref{fig:profiles2} we display results for a case where $\e^{nl}$ is set to be much lower than for the cases in Fig.~\ref{fig:profiles}. This corresponds to the nanoparticles having a fairly poor solubility in the liquid. As a consequence, we see from the density profiles that the nanoparticles gather to form a dense clump with only a relatively small amount of liquid within. Interestingly, the density of the liquid is highest around the edge of the droplet. We present this particular result to illustrate that the density profiles are not always as simple as those presented in Fig.~\ref{fig:profiles}. Before discussing any further DFT results and in particular how the droplet size varies, depending on the number of nanoparticles within and the value of the parameter $\e^{nl}$, we first present our capillarity-approximation based thermodynamic model in Sec.~\ref{sec:cap_model} below. Our aim is to compare results from the microscopic DFT model with this mesoscopic capillarity-approximation model.  We show in the following section that the two are in excellent agreement.

\section{Capillarity model}
\label{sec:cap_model}

Since our DFT calculations in the previous section are for a 2D system, here we present our capillarity approximation (CA) model in 2D. However, it is no more difficult to construct it instead in 3D, so at various points in the following we additionally give the corresponding 3D equations.

Consider first a system of volume $V$ (strictly, area in 2D) containing just the bulk vapour phase. The grand potential of the system is then just
\begin{equation}
\Omega_{\rm vap}=-p_{\rm vap}V,
\label{eq:Om_g}
\end{equation}
where $p_{\rm vap}$ is the bulk pressure in the vapour phase. In our model we assume this is given by Eq.~\eqref{eq:p} in the limit $\rho^n\to0$. Note that the vapour density $\rho^l$ is determined by the chemical potential $\mu$, which we assume to be specified. Alternatively, one could assume that the vapour density (i.e.\ the humidity) is given and then the chemical potential can be calculated via Eq.~\eqref{eq:mu}, with $\rho^n\to0$.

Now consider the case when the system also contains a circular (in 2D) droplet of radius $R$, surrounded by the vapour phase; see e.g.\ Fig.~\ref{fig:diagram}. The grand potential can now be approximated as
\begin{equation}
\Omega_{\rm drop}=-p_{\rm drop}\pi R^2-p_{\rm vap}(V-\pi R^2)+\gamma 2\pi R.
\end{equation}
The first term is the bulk contribution due to the droplet which has volume (area in 2D) equal to $\pi R^2$; c.f.\ Eq.~\eqref{eq:Om_g}. The second term is the corresponding contribution due to the vapour filling the remainder of the system. The final term is the contribution from the interface between the liquid and gas phases; $\gamma$ is the interfacial tension. The quantity of relevance in the calculations that follow is the difference between these, $\Delta\Omega=\Omega_{\rm drop}-\Omega_{\rm vap}$, which is given by
\begin{equation}
\Delta\Omega=-(p_{\rm drop}-p_{\rm vap})\pi R^2+\gamma 2\pi R.
\label{eq:D_Om_2D}
\end{equation}
In 3D, the above should be replaced by the following
\begin{equation}
\Delta\Omega_{3D}=-(p_{\rm drop}-p_{\rm vap})\frac{4}{3}\pi R^3+\gamma 4\pi R^2.
\label{eq:D_Om_3D}
\end{equation}
Here we assume that the interfacial tension $\gamma$ is that of the pure vapour-liquid system at the given temperature. For our model, when $\beta\e^{ll}=1.2$, we obtain $\beta\gamma\sigma=0.68587$, which is calculated in the usual way -- see e.g.\ \cite{evans1979nature, hughes2014introduction}. The fixed number of nanoparticles in the droplet means that the nanoparticle density in the droplet is (in 2D)
\begin{equation}
\rho_n=\frac{N_n}{\pi R^2},
\label{eq:rho_n}
\end{equation}
or (in 3D)
\begin{equation}
\rho_n=\frac{3N_n}{4 \pi R^3}.
\label{eq:rho_n_3D}
\end{equation}
Substituting Eq.~\eqref{eq:rho_n} into Eq.~\eqref{eq:p} [together with Eq.~\eqref{eq:f_bulk}], we obtain the following expression for the pressure in the drop (in 2D)
\begin{align}
\label{eq:p_drop}
p_{\rm drop}(\rho^l,R)= & k_BT\left[\rho^l \ln{\rho^l} + \left(1 - \rho^l - \frac{N_n}{\pi R^2} \right) \ln\left(1 - \rho^l - \frac{N_n}{\pi R^2}\right) + \frac{N_n}{\pi R^2} \ln{\frac{N_n}{\pi R^2}}\right]
\nonumber
\\ 
& - 3 \e^{ll}(\rho^l)^2 - 6 \e^{nl} \rho^l \left(\frac{N_n}{\pi R^2}\right) - 3 \e^{nn} \left(\frac{N_n}{\pi R^2}\right)^2.
\end{align}
Substituting this into Eq.~\eqref{eq:D_Om_2D}, we obtain an expression for $\Delta\Omega=\Delta\Omega(\rho^l,R)$, that is a function of $(\rho^l,R)$. The equilibrium droplet radius $R$ is the value that minimises the free energy, therefore we require
\begin{equation}
\frac{\partial \Delta\Omega(\rho_l,R)}{\partial R}=0.
\label{eq:min_DOmega}
\end{equation}
Similarly, substituting Eq.~\eqref{eq:rho_n} into Eq.~\eqref{eq:mu} yields the following second equation
\begin{equation}
k_BT\ln(\rho^l) - k_BT\ln\left(1 - \rho^l - \frac{N_n}{\pi R^2}\right) - 6\e^{ll}\rho^l - 6\e^{nl}\left(\frac{N_n}{\pi R^2}\right) - \mu=0,
\label{eq:2nd}
\end{equation}
that is also a function of the two unknowns $(\rho^l,R)$. This equation corresponds to requiring that the liquid density in the drop equals the higher of the two possible values determined by the selected value of the chemical potential $\mu$. In other words, we require the value of the chemical potential within the droplet to be the same as that in the surrounding vapour. We then simply solve numerically (using fsolve in Maple) the pair of simultanous Eqs.~\eqref{eq:min_DOmega} and \eqref{eq:2nd} for the two unknowns, $\rho^l$ and $R$. From these, we can then easily obtain the total amount of liquid in the droplet as (in 2D)
\begin{equation}
\Gamma_l=\rho^l\pi R^2,
\label{eq:V_liq}
\end{equation}
or (in 3D)
\begin{equation}
\Gamma_l=\rho^l\frac{4}{3}\pi R^3.
\label{eq:V_liq_3D}
\end{equation}

\begin{figure}[t]
   \centering
   \includegraphics[width=4in]{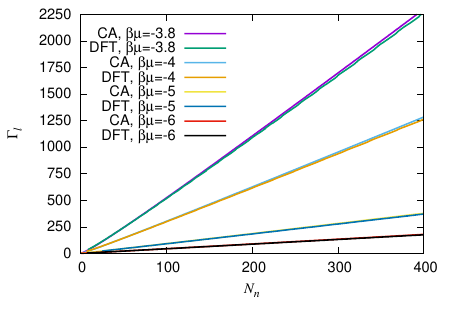}
   \caption{Plots of the amount of liquid $\Gamma_l$ in the droplet as a function of the number of nanoparticles in the droplet $N_n$, for various values of the chemical potential $\mu$, i.e.\ for varying humidity. We compare results from the capillarity approximation (CA), where $\Gamma_l$ is calculated via Eq.~\eqref{eq:V_liq}, with those from DFT, where $\Gamma_l$ is calculated via Eq.~\eqref{eq:DFT_Gamma}. The results here are for $\beta\e^{ll}=1.2$, $\beta\e^{nn}=0.9$, and $\beta\e^{nl}=1.5$. Note that $\beta\mu=-3.8$ corresponds to to a relative humidity of $H_r=80\%$; $\beta\mu=-4$ corresponds to $H_r=65\%$; $\beta\mu=-5$ corresponds to $H_r=23\%$; and $\beta\mu=-6$ corresponds to $H_r=8\%$. Note also that the CA lines for $\beta\mu=-5$ and $\beta\mu=-6$ are barely visible, since the corresponding DFT result is almost identical.}
   \label{fig:ads_Enl1_5}
\end{figure}

\begin{figure}[t] %  figure placement: here, top, bottom, or page
   \centering
   \includegraphics[width=4in]{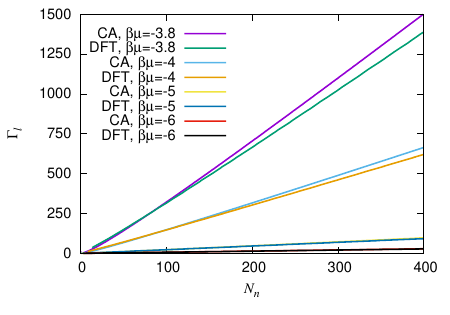}
   \caption{The results here are the same as those in Fig.~\ref{fig:ads_Enl1_5}, except now the strength of the attraction between the nanoparticles and the liquid is much lower, with $\beta\e^{nl}=1$; i.e.\ these are results for nanoparticles with a much lower solubility. Note the change in the range of the vertical axis compared to Fig.~\ref{fig:ads_Enl1_5}. Also, note again that the CA lines for $\beta\mu=-5$ and $\beta\mu=-6$ are barely visible, since the corresponding DFT result is almost identical.}
   \label{fig:ads_Enl1}
\end{figure}

In Fig.~\ref{fig:ads_Enl1_5} we compare the results from this simple CA with results from the DFT for the case when $\beta\e^{ll}=1.2$, $\beta\e^{nn}=0.9$, $\beta\e^{nl}=1.5$ and for various values of the chemical potential $\mu$ (i.e.\ for various values of the humidity). To compare, we plot $\Gamma_l$ from the CA in Eq.~\eqref{eq:V_liq} together with the result from the DFT, where we determine the amount of liquid in the droplet via the following sum over lattice sites within the droplet
\begin{equation}
\Gamma_l=\sum_{\rho_\ii^l>0.99\rho_{\rm max}^l}\rho_\ii^l,
\label{eq:DFT_Gamma}
\end{equation}
where $\rho_{\rm max}^l=\sup(\{\rho^l_\ii\})$ is the maximum liquid density in the system (i.e.\ in the centre of the drop). Thus, we do not include the material in the surrounding vapour, when determining the mass of liquid in the droplet. Choosing the threshold as $\rho_\ii^l>0.99\rho_{\rm max}^l$ is somewhat arbitrary. However, changing this to e.g.\ $\rho_\ii^l>0.9\rho_{\rm max}^l$ only slightly changes our results. From the results in Fig.~\ref{fig:ads_Enl1_5}, we see that the agreement between the CA model and the DFT is rather good. We believe the main source of error in our CA model is in our decision to use the value for the interfacial tension $\gamma$ obtained for the pure liquid system. In reality, the presence of the nanoparticles changes the value of $\gamma$. However, given the agreement we see in Fig.~\ref{fig:ads_Enl1_5}, we conclude that, at least for the present system, this is a reasonable approximation to make.

In Fig.~\ref{fig:ads_Enl1} we show results for the case where $\beta\e^{nl}=1$ (all other parameters are the same as in Fig.~\ref{fig:ads_Enl1_5}). This corresponds to a much lower value for the strength of attraction between the nanoparticles and the liquid and therefore corresponds to nanoparticles which have a much lower solubility. We see that as a result of the weaker attraction between the nanoparticles and the liquid, the droplets are therefore smaller, as one should expect. Note again the good agreement between the CA and the DFT. Thus, we conclude the simple CA is indeed able to capture the effects of varying attraction strengths on determining the equilibrium droplet size.

\section{Dynamics}
\label{sec:dynamics}

We assume that the non-equilibrium dynamics is governed by dynamical density functional theory (DDFT), as described in detail in~\cite{chalmers2017dynamical}. We could include hydrodynamics, e.g.\ following \cite{areshi2019kinetic}, but here restrict ourselves to the usual overdamped DDFT. DDFT has been derived for Brownian particles suspended in a liquid~\cite{marconi1999dynamic, marconi2000dynamic}, such as the nanoparticles studied here, and has also been developed for molecular liquids~\cite{archer2006dynamical, archer2009dynamical}. The latter case is particularly accurate when the fluid is close to equilibrium, which is the case here. As such, DDFT is applicable to both components of the system studied here.

The (lattice) DDFT for the two component system consists of a coupled pair of equations at each lattice site:
\begin{align}
    \frac{\partial \rho_\ii^l}{\partial t} &= \nabla \cdot 
    \Big[ M_l\rho_\ii^l \nabla \frac{\partial F}{\partial \rho_\ii^l}
    \Big] \label{eq:DDFTl}\\
    \frac{\partial \rho_\ii^n}{\partial t} &= \nabla \cdot 
    \Big[ M_n\rho_\ii^n \nabla \frac{\partial F}{\partial \rho_\ii^n}
    \Big]. \label{eq:DDFTn}
\end{align}
Note that in contrast to the DDFT presented in~\cite{chalmers2017dynamical} we have constant mobility coefficients $M_l$ and $M_n$, for both the liquid and nanoparticles.  It is simple to extend the DDFT to the non-constant case, but we do not find it necessary here. Note also that we set $M_l=M_n=1$ for simplicity.

The ensemble-averaged densities at site $\ii$, $\rho_\ii^l$ and $\rho_\ii^n$ are now functions of time. The differential operators in~\eqref{eq:DDFTl} and~\eqref{eq:DDFTn} are finite difference approximations on the lattice.  Care is needed when applying these to prevent numerical instabilities.  We take the approach detailed in~\cite{chalmers2017dynamical}, in particular alternating the direction of the spatial finite difference.  Here we apply this method in both spatial directions, in contrast to~\cite{chalmers2017dynamical} where this was only required in the direction parallel to their wall.  For the time evolution we use an Euler scheme but note that the results are almost indistinguishable from those using higher order schemes such as the fourth order adaptive Runge-Kutta scheme implemented in Matlab's ode45~\cite{dormand1980family}.

\begin{figure}
\begin{tabular}{l@{\hskip 3mm} c@{\hskip 5mm}|@{\hskip 5mm}c}

& {\large $\beta \mu = -3.8$} & {\large $\beta \mu = -5$} \\

\raisebox{15mm}{\rotatebox{90}{\large $t=0$}} &
\includegraphics[width=0.45\textwidth]{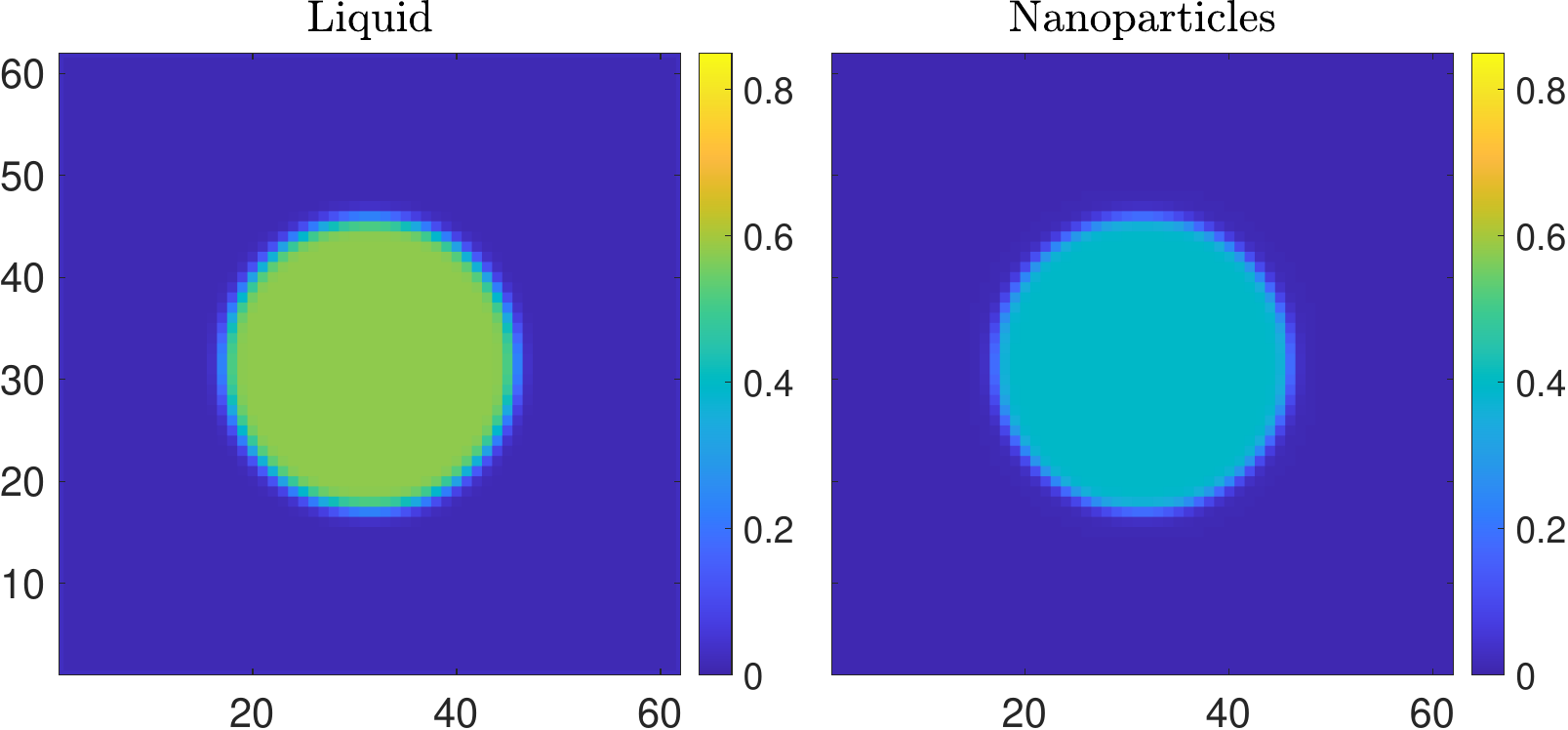}&
\includegraphics[width=0.45\textwidth]{001.pdf}\\

\raisebox{10mm}{\rotatebox{90}{\large $t=4,000$}} &
\includegraphics[width=0.45\textwidth]{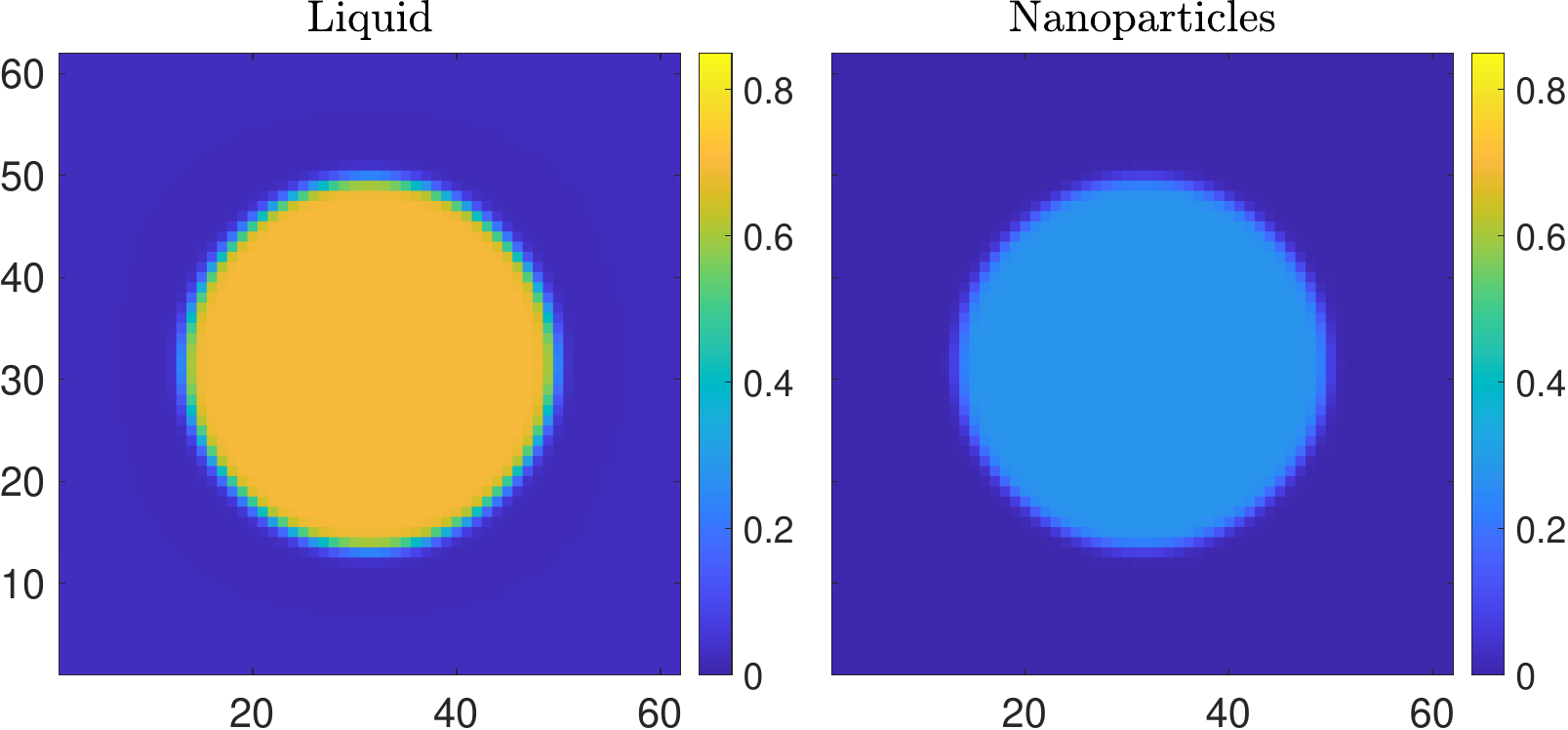}&
\includegraphics[width=0.45\textwidth]{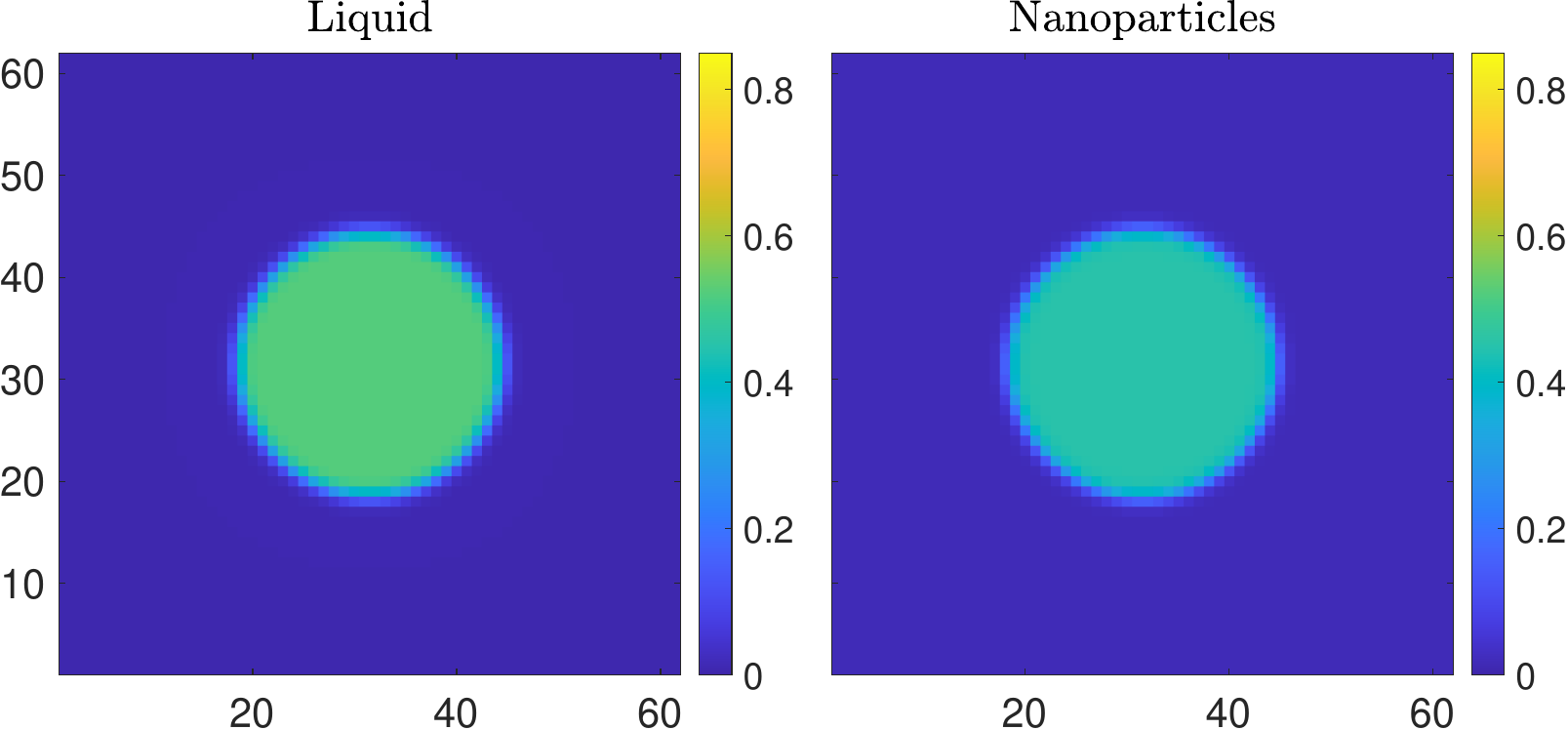}\\

\raisebox{8mm}{\rotatebox{90}{\large $t=20,000$}} &
\includegraphics[width=0.45\textwidth]{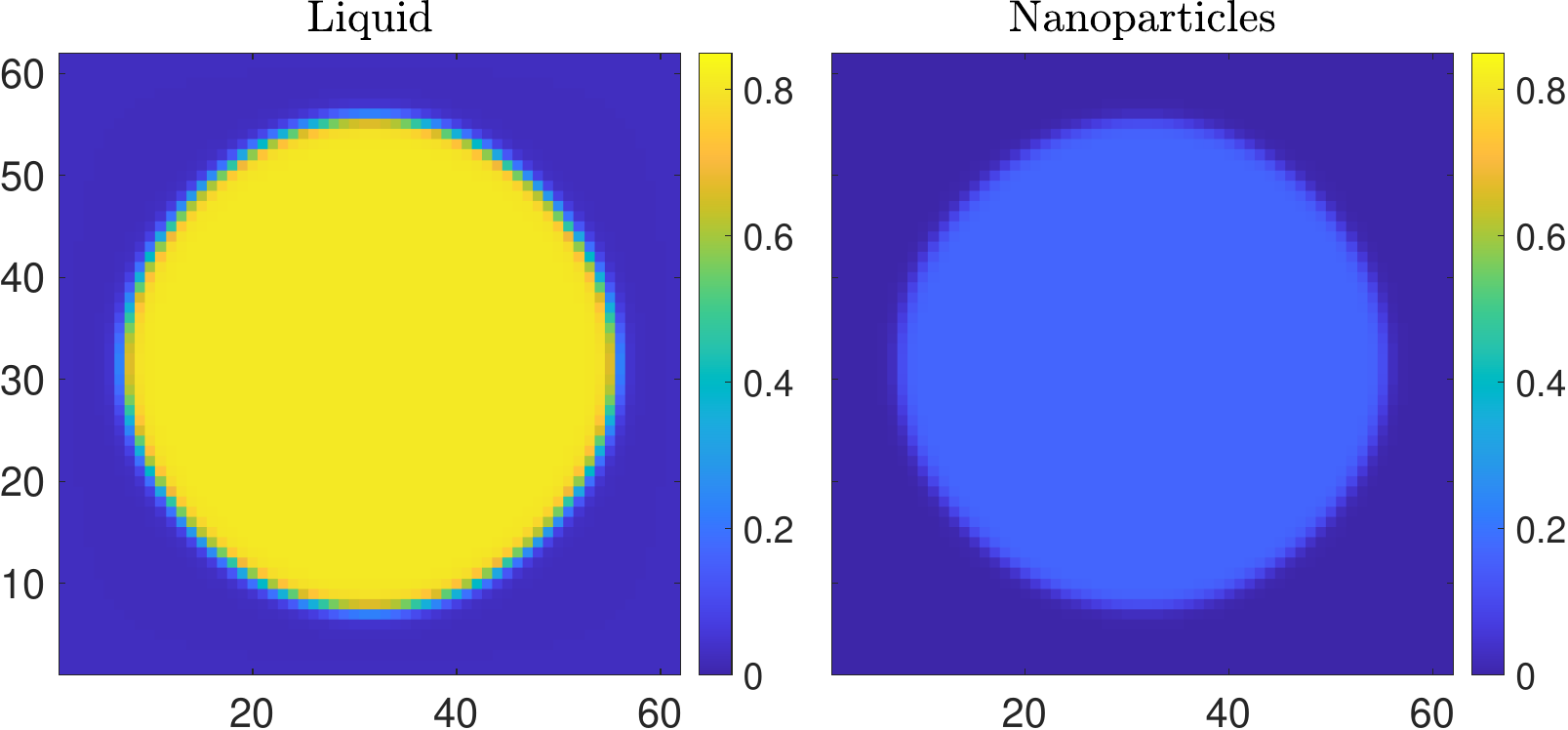}&
\includegraphics[width=0.45\textwidth]{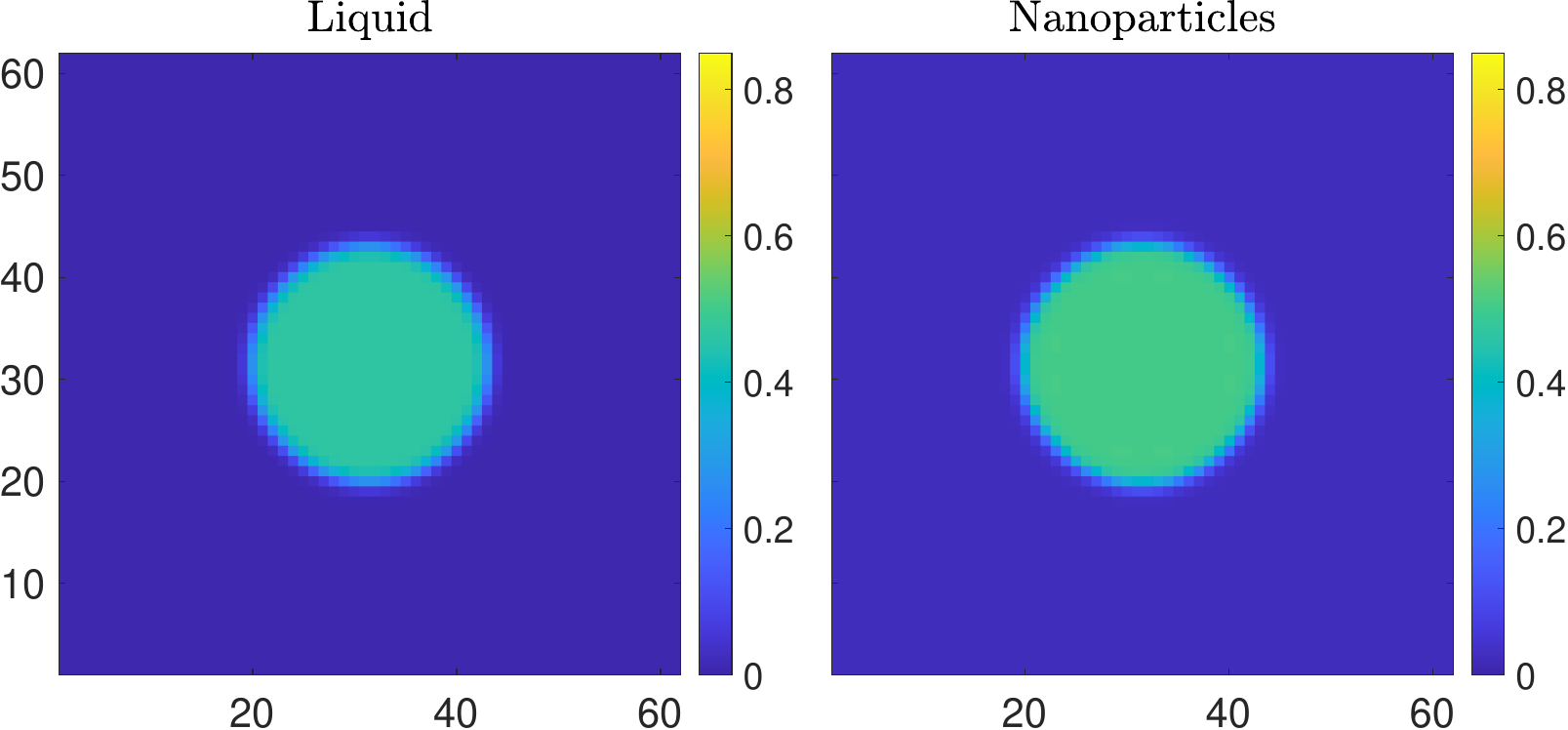}
\end{tabular}

\caption{Snapshots of DDFT simulations, starting from the DFT equilibrium with $\beta\mu = -4.5$ at times $t=0$, $t=4000$, and $t=20000$ (top to bottom).
Pairs of plots on the left (right) show the DDFT dynamics for the liquid and the nanoparticles when the liquid density on the boundary of the box is set to the value given by the equilibrium DFT computations with $\beta\mu = -3.8$ ($\beta\mu = -5$).  As expected from the equilibrium calculations, the droplet on the left (right) grows (shrinks) over time.}
\label{fig:DDFTSnapshots}

\end{figure}

\begin{figure}
\includegraphics[width=0.8\textwidth]{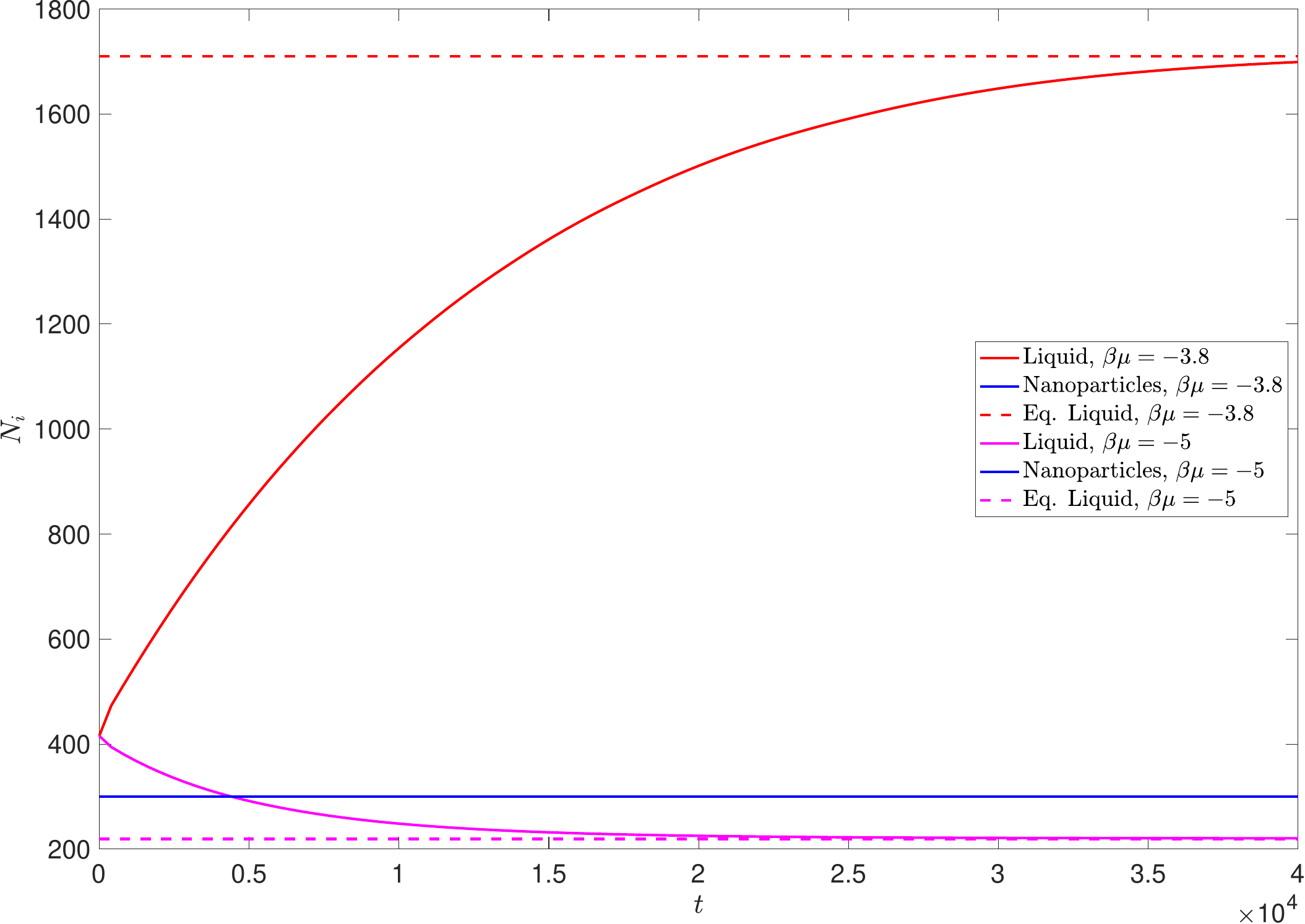}
\caption{The mass evolution of the liquid and nanoparticles over time for the DDFT simulations shown in Fig.~\ref{fig:DDFTSnapshots}.  Dashed lines denote the liquid masses in the corresponding DFT computations.  Note, in particular, that increasing (decreasing) $\beta\mu$ from the value of $-5.5$ which provides the initial condition causes the amount of liquid in the box to decrease (increase).}
\label{fig:DDFTMasses}

\end{figure}

In Figs.~\ref{fig:DDFTSnapshots} and~\ref{fig:DDFTMasses} we show the results for two DDFT simulations.  Both start from the DFT equilibrium droplet for $\beta\e^{ll}=1.2$, $\beta\e^{nn}=0.9$, and $\beta\e^{nl}=1.5$ with $\beta\mu = -5.5$ (also displayed in Fig.~\ref{fig:profiles}).  To induce dynamics, we set the liquid density on the boundary of the box to a constant value, corresponding to that of a DFT simulation with a different value of $\mu$.  This results in the DDFT simulation equilibrating to the corresponding DFT result.  We show two cases, $\beta\mu = -3.8$ and $\beta\mu=-5$.  The former case, displayed in the two left hand columns of Fig.~\ref{fig:DDFTSnapshots}, corresponds to a droplet having moved into a more humid environment than that in which it was initially formed. This results in an increase in the amount of liquid in the box, as additional liquid diffuses into the box from the boundary and then condenses onto the droplet, making it grow in size. In contrast, the DDFT results in the two right hand columns of Fig.~\ref{fig:DDFTSnapshots} correspond to the droplet moving to an environment that has a lower humidity than that where it was formed. This is the typical case of aerosol droplets that form in a person's mouth or respiratory system at a temperature of about 37$^\circ$C at 100\% relative humidity that then subsequently move out of the mouth into the air, which is a less humid environment. The DDFT results show the size of the droplet decreasing over time, as it reduces to a smaller equilibrium size. Note, however, that the droplet never completely evaporates; the nanoparticles within mean that it remains stable, albeit at a smaller size.  This can also be seen in Fig.~\ref{fig:DDFTMasses}, where we plot the total mass of liquid in the simulation box as a function of time for these two cases.  The relative errors (in the $\ell_1$ norm) between the final dynamic profiles and the corresponding equlibria are below 1\% in all cases after time $4 \times 10^4$; this can be further reduced by running the DDFT simulations for longer times.

We note that a crucial aspect of these DDFT simulations relates to the choice of boundary conditions.  In the examples presented in Figs.~\ref{fig:DDFTSnapshots} and~\ref{fig:DDFTMasses}, the nanoparticles are largely concentrated in the middle of the simulation domain and the interparticle attraction prevents any significant diffusion away from this area and, as such, periodic boundary conditions for the nanoparticles are an appropriate choice (any other sensible boundary condition will give almost indistinguishable results).  However, the boundary conditions used for the liquid are much more important. In particular, the distance from the surface of the droplet to the edge of the boundary where we keep the liquid density set to the value corresponding to the desired chemical potential value (i.e.\ relative humidity value, specified for each simulation), is all-important. If the box size is increased, i.e.\ the distance from the droplet to the boundary is increased, then the DDFT simulations take correspondingly longer to equilibrate. What determines the time for a droplet to equilibrate is a combination of two processes: the first relates to the time it takes for liquid to move out of the droplet, across the liquid-vapour interface. The second part of the process is that of the liquid diffusing through the vapour surrounding the droplet, to reach the boundary and be absorbed. The second process is well understood: for free diffusion from the centre to the edge of a circular domain, the total amount in the system $N_l(t)$, given by Eq.~\eqref{eq:N_l}, follows the well-known result
\begin{equation}
N_l(t)\approx[N_l(0)-N_l(t\to\infty)]e^{-\lambda t} +N_l(t\to\infty),
\label{eq:exponential}
\end{equation}
i.e.\ the amount of liquid decreases exponentially over time with the rate constant $\lambda=D(j_{1,0}/a)^2$, where $D=M_lk_BT$ is the diffusion coefficient, $j_{1,0}$ is the first zero of the Bessel $J_0(x)$ function, and $a$ is the radius of the domain. This is the situation our model reduces to in the limit where the densities of the liquid and nanoparticles are small everywhere (where the DDFT equations \eqref{eq:DDFTl} and \eqref{eq:DDFTn} reduce to diffusion equations). However, for the cases of interest here, the additional process of particles crossing the liquid-vapour interface makes the whole equilibration process much slower. In our DDFT simulations we still observe $N_l(t)$ varying over time with the simple exponential decay form in Eq.~\eqref{eq:exponential} (see Fig.~\ref{fig:DDFTMasses}), but the rate constant $\lambda$ that we observe is much smaller than the result quoted above for the case for simple diffusion from the centre to the edge of the domain, due to the additional interface crossing process. Nonetheless, these considerations demonstrate why the distance from the droplet to the edge of the simulation box (i.e.\ the size we assume for the diffusive boundary layer around the droplets) is important for determining the overall time scale of the equilibration process. That said, we find that as long as the edge of droplet is $\approx5$ or more lattice sites away from the boundary of the box, then snapshots over time from simulations in a small box and a larger box are almost indistinguishable. 

\begin{figure}

\includegraphics[width=0.48\textwidth]{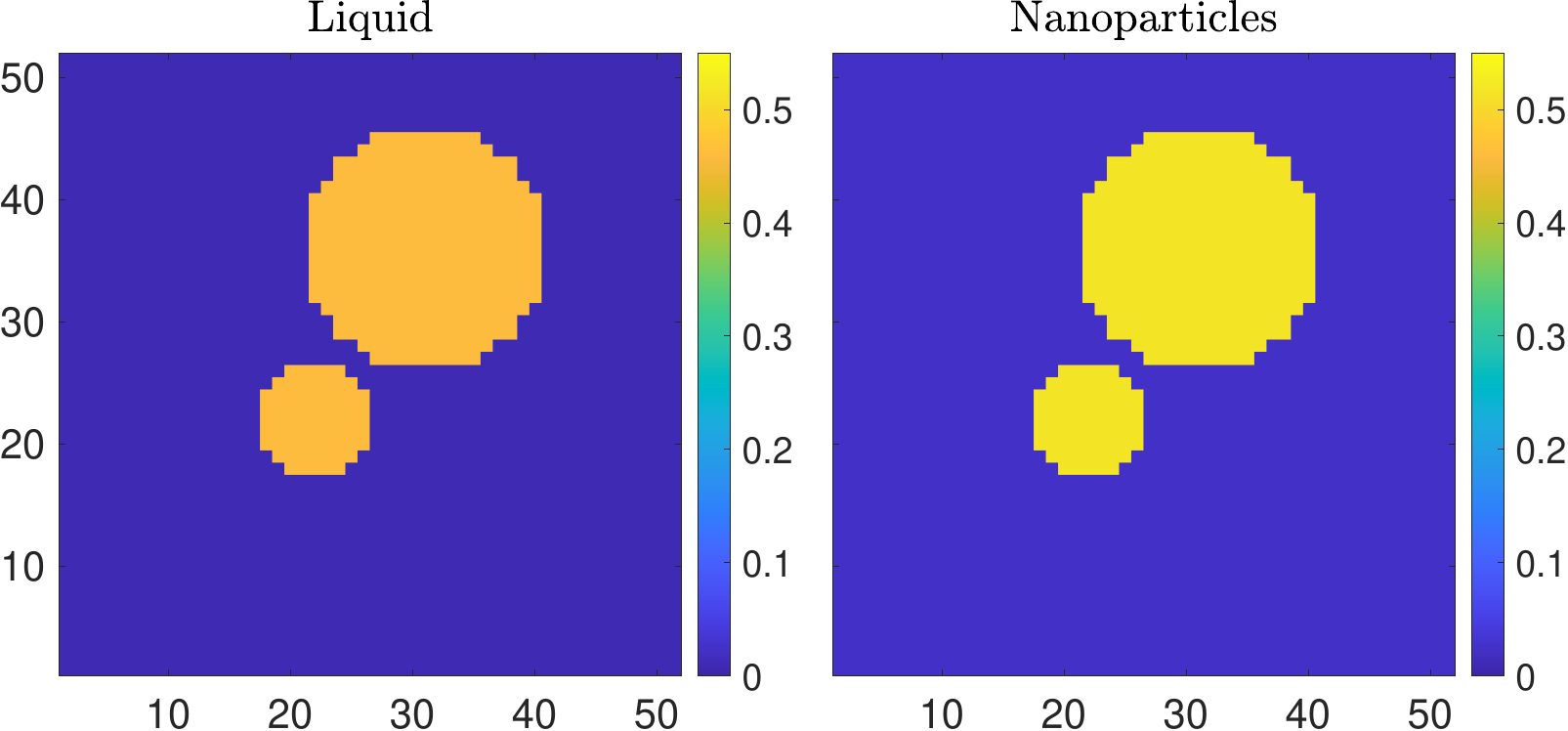} % t = 0
\includegraphics[width=0.48\textwidth]{Merging_001.pdf} % t = 0

\includegraphics[width=0.48\textwidth]{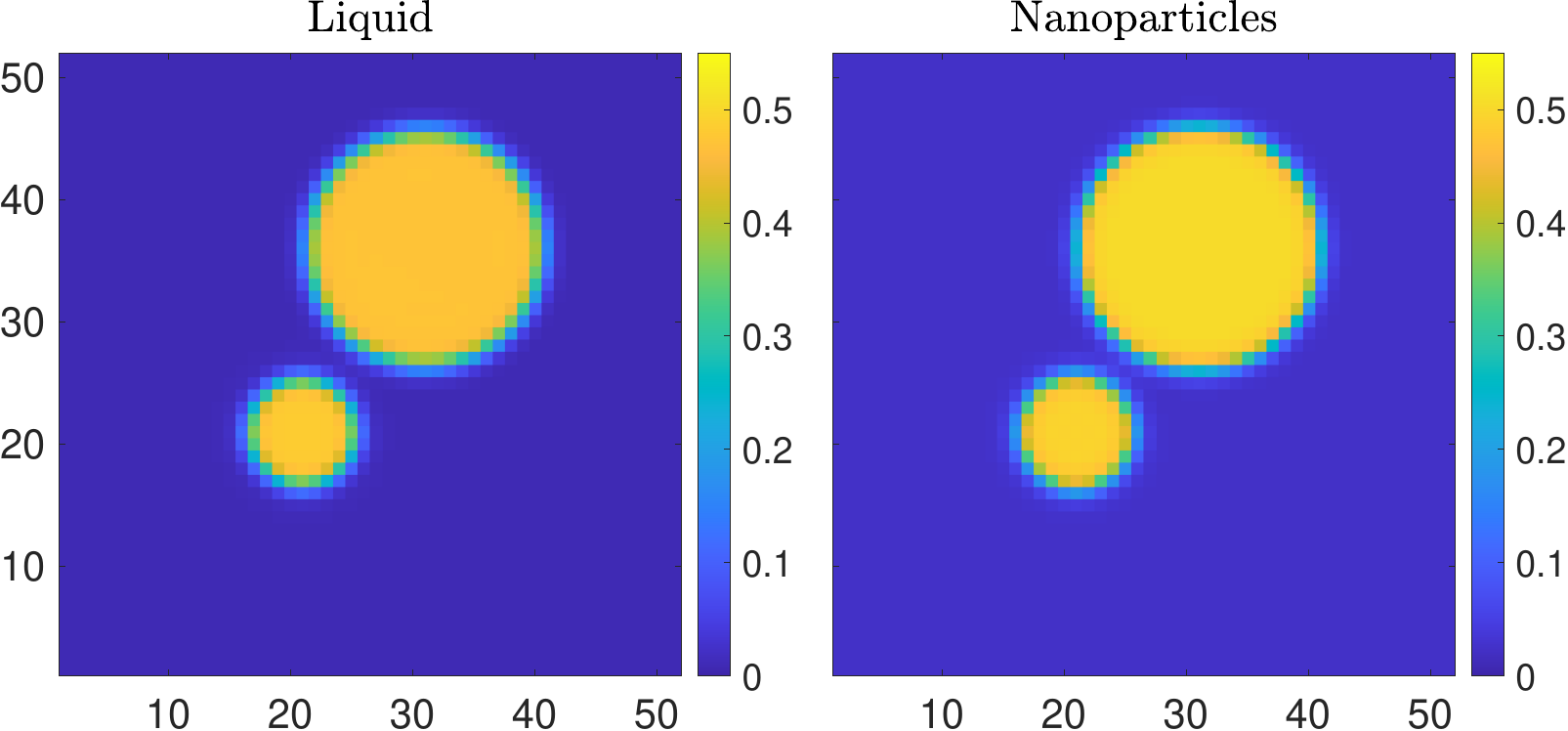} % t = 20
\includegraphics[width=0.48\textwidth]{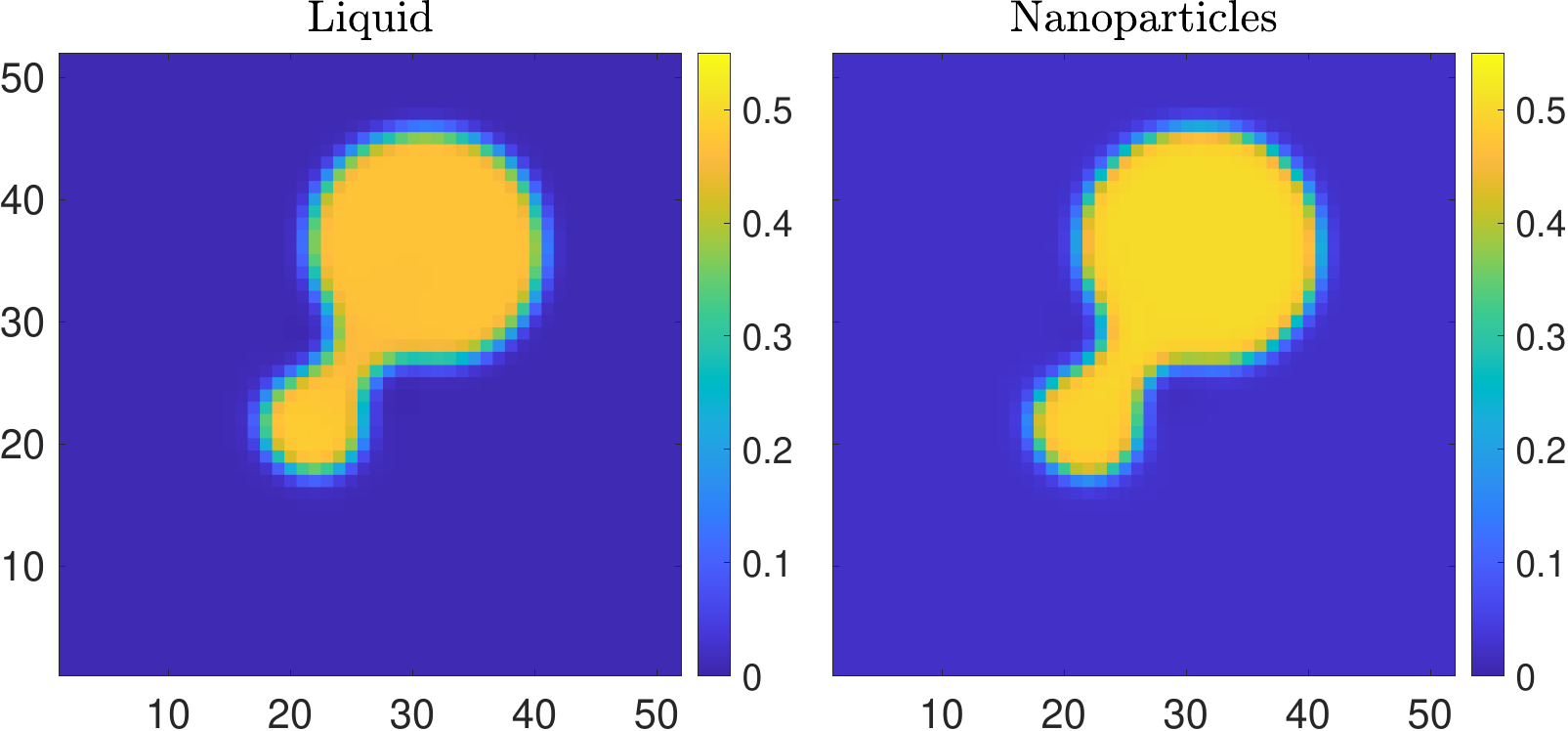} % t = 10

\includegraphics[width=0.48\textwidth]{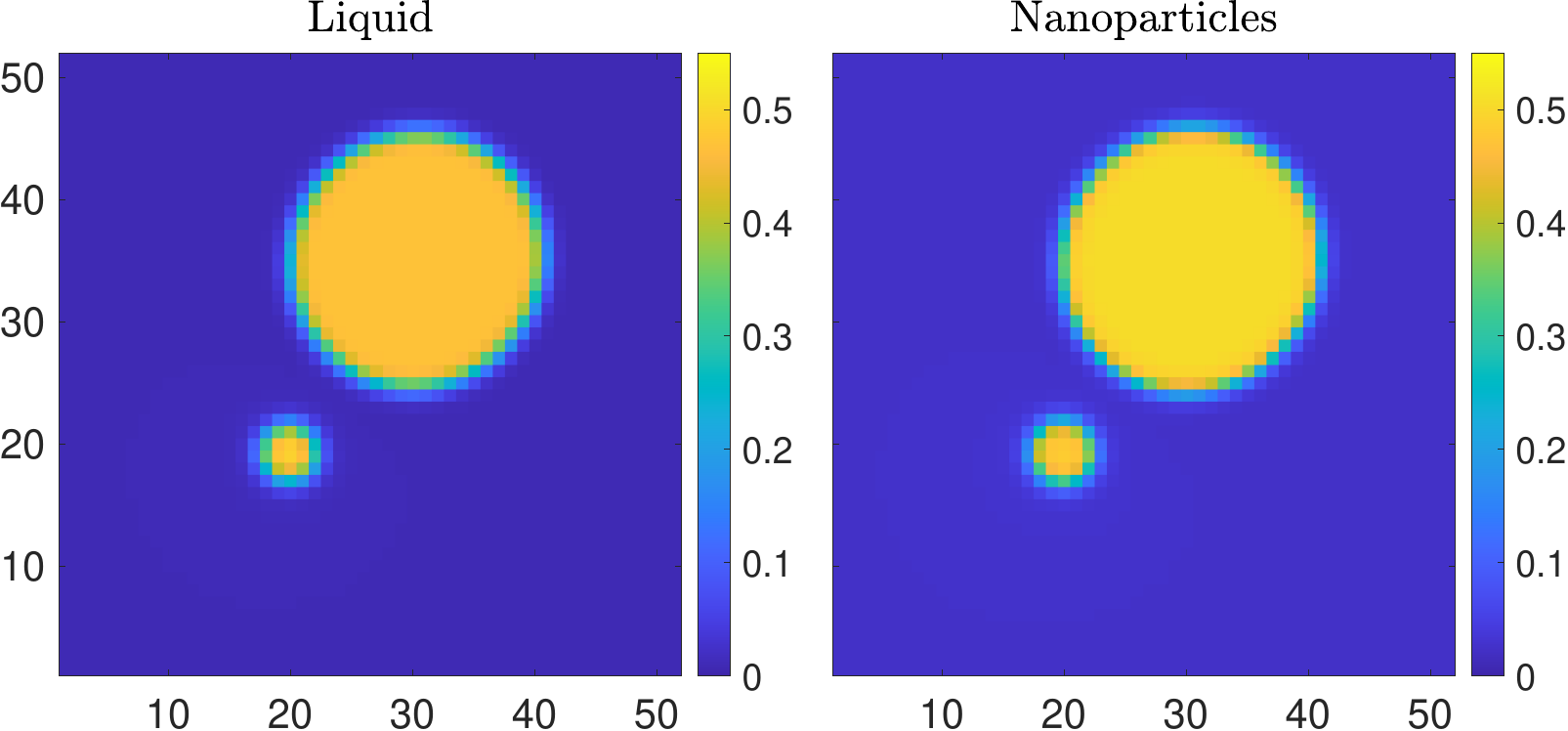} % t = 1200
\includegraphics[width=0.48\textwidth]{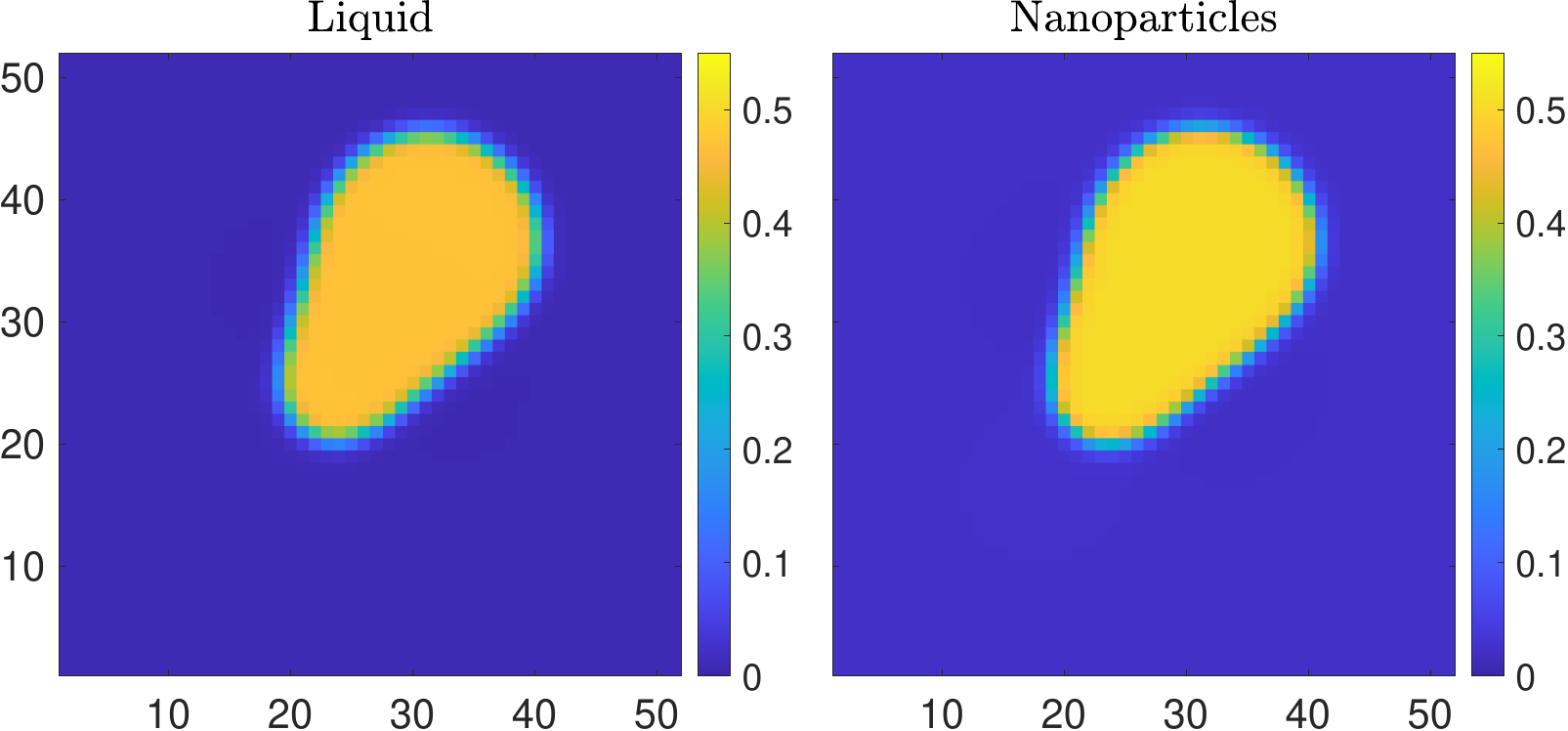} % t = 100

\includegraphics[width=0.48\textwidth]{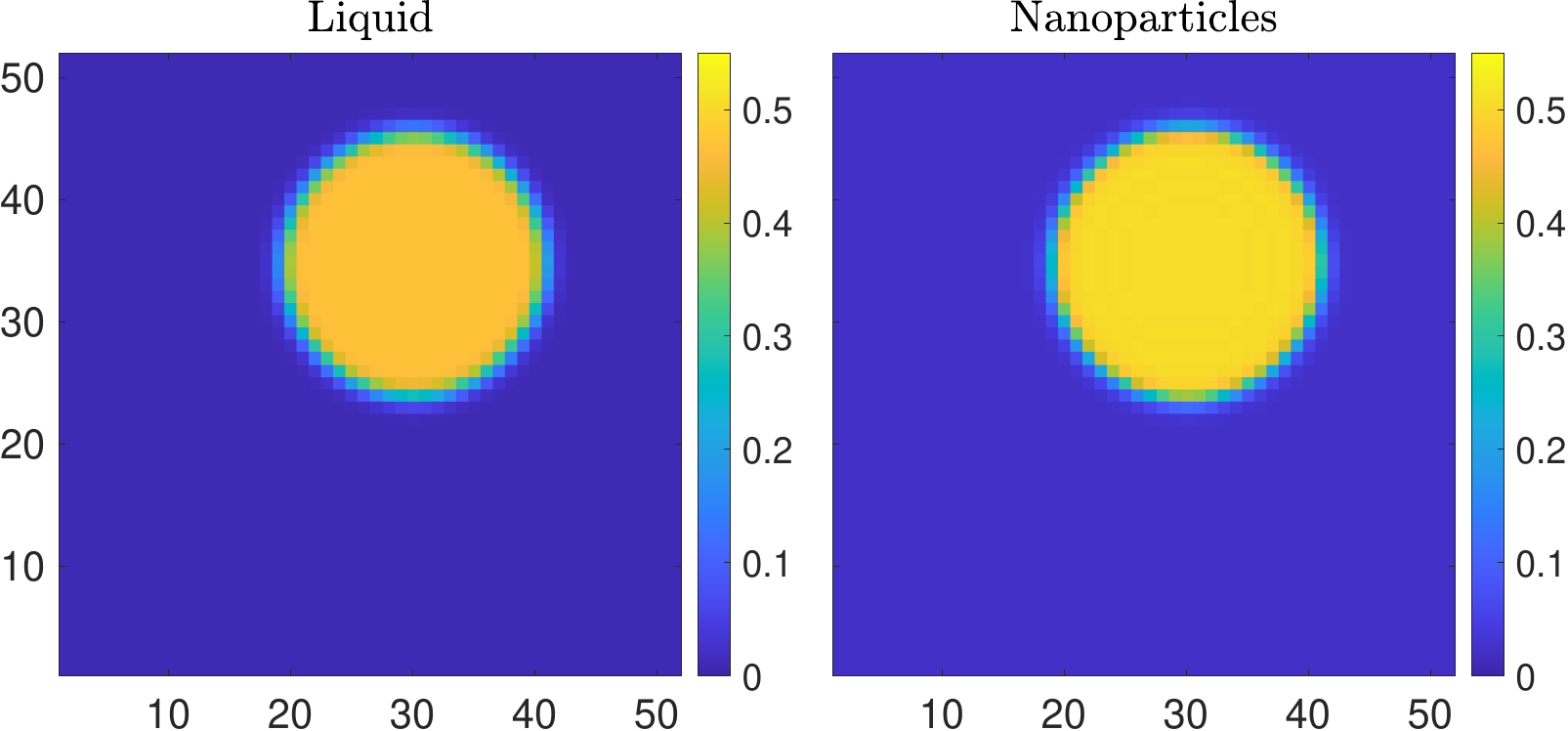} % t = 1000
\includegraphics[width=0.48\textwidth]{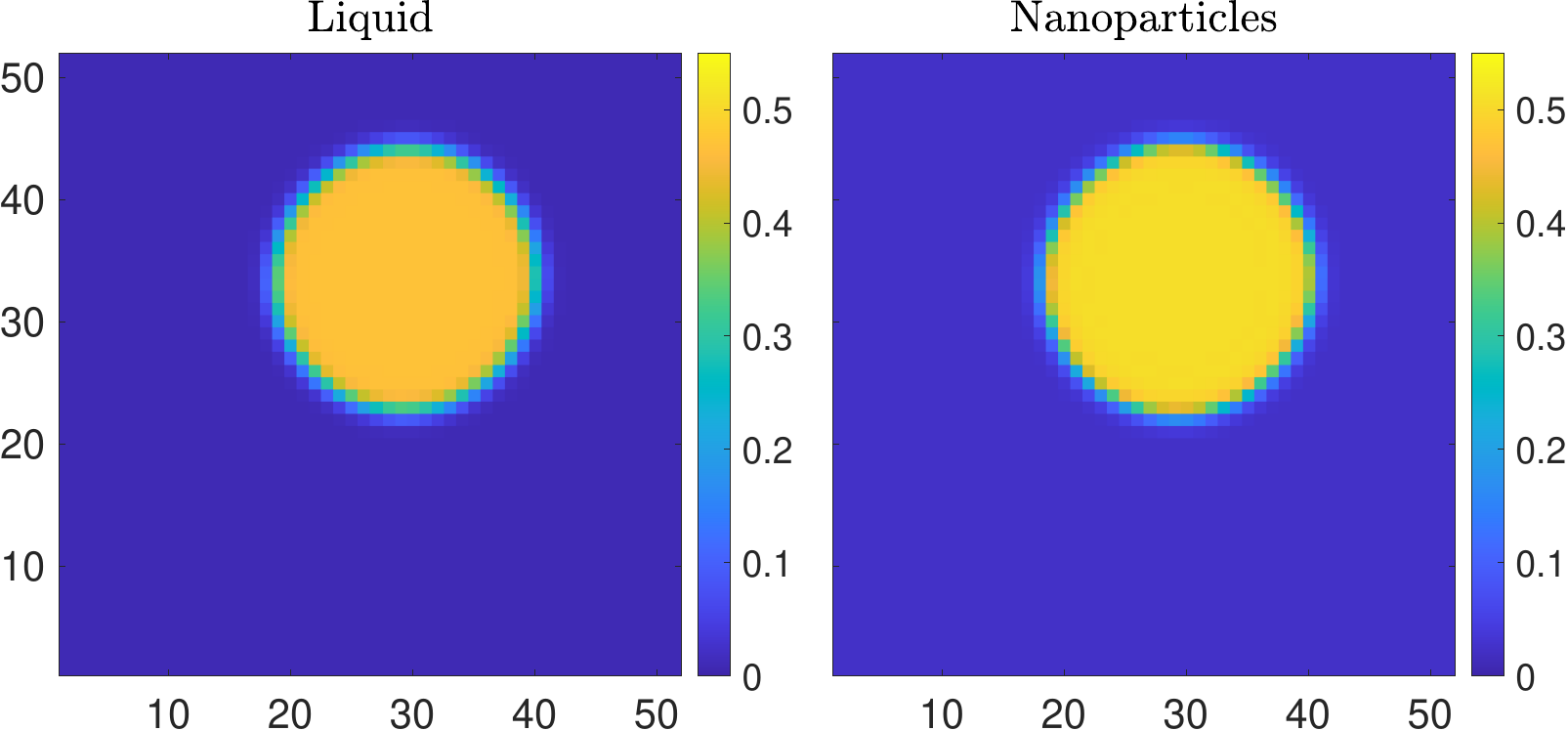} % t = 2000

\caption{Snapshots from DDFT simulations, starting from two circular distributions at times $t=0$, $t=20$, $t=1200$, and $t=2000$ (left, top to bottom) and $t=0$, $t=10$, $t=100$, and $t=1000$ (right, top to bottom).  The simulations differ only in the initial location of the smaller droplet (see text for details). However, this small difference in initial condition makes a very significant difference to the manner in which the two droplets coalesce.}
\label{fig:DDFTSnapshots_coalesence}

\end{figure}

Our DDFT model can be used to predict the dynamics of aerosol droplets in a great variety of different situations. For example, if we included the external potentials $\Phi_\ii^l$ and $\Phi_\ii^n$ due to a surface, we could model the slow impact of droplets with a surface and the subsequent spreading and drying process. Illustrative results for the later part of this dynamics can be found e.g.\ in Refs.~\cite{chalmers2017dynamical, perez2021changing}. Here, we restrict ourselves to presenting a pair of illustrative results corresponding to the coalescence of two different sized droplets. In Fig.~\ref{fig:DDFTSnapshots_coalesence} we show results for droplets joining for the case when $\beta\e^{ll}=1.2$, $\beta\e^{nn}=0.9$, $\beta\e^{nl}=1.5$ and $\beta\mu = -5$. The initial conditions correspond simply to setting all of the lattice sites within two circular regions to the density values at the centre a single equilibrium droplet for this set of parameter values, while the density outside the circles is set to be that of the corresponding vapour in the single-droplet DFT calculation. The radii of the two circles (i.e.\ initial radii of the two droplets) are 5 and 10 with centres at (20,20) and (30,35) for the left hand simulations and (21,21) and (30,35) for the right hand simulations in Fig.~\ref{fig:DDFTSnapshots_coalesence}. Thus, the only difference between the two simulations is that the smaller droplet is moved sightly closer to the bigger droplet for the right hand set of results. We see however that this very small change makes a big difference to the dynamics. For the case on the left, the small droplet shrinks and joins the larger droplet via diffusion through the vapour, while in the case on the right the whole droplet moves and joins the larger one. Why it is that one sees one process at one distance and the other at a slightly different distance was studied in detail in Ref.~\cite{pototsky2014coarsening} in the context of a different DDFT. The mechanism followed by the case on the left is termed joining via the Ostwald mode, which was first described in Refs.~\cite{lifshitz1961kinetics, wagner1961theorie} to understand the process of Ostwald ripening, while the mechanism followed by the case on the right is termed the translation mode. One can calculate which mode will dominate by linearising the DDFT equation around the initial state and then one obtains two distinct eigenfunctions corresponding to each of these modes. The mode one observes actually occurring is the one with the largest corresponding eigenvalue \cite{pototsky2014coarsening}. These results illustrate just one possibility in the hugely complex dynamics of aerosol droplets. Note also that our DDFT model assumes an over-damped (diffusive) dynamics. If we extended our theory to include the effects of inertia, then, for example, the evaporation and coalescence of droplets in the turbulent airflow following a person sneezing could be investigated \cite{renzi2020life}. However, we do not pursue that direction here. The examples in Fig.~\ref{fig:DDFTSnapshots_coalesence} illustrate that the interplay of dynamics with an underlying complex free energy landscape can result in rather complicated dynamics. In the following section, we illustrate this point further, albeit with examples that correspond to a somewhat unlikely (in Nature) initial state.

\section{Comparing DDFT to Picard}
\label{sec:DDFT_v_Picard}

\begin{figure}
\includegraphics[width=0.45\textwidth]{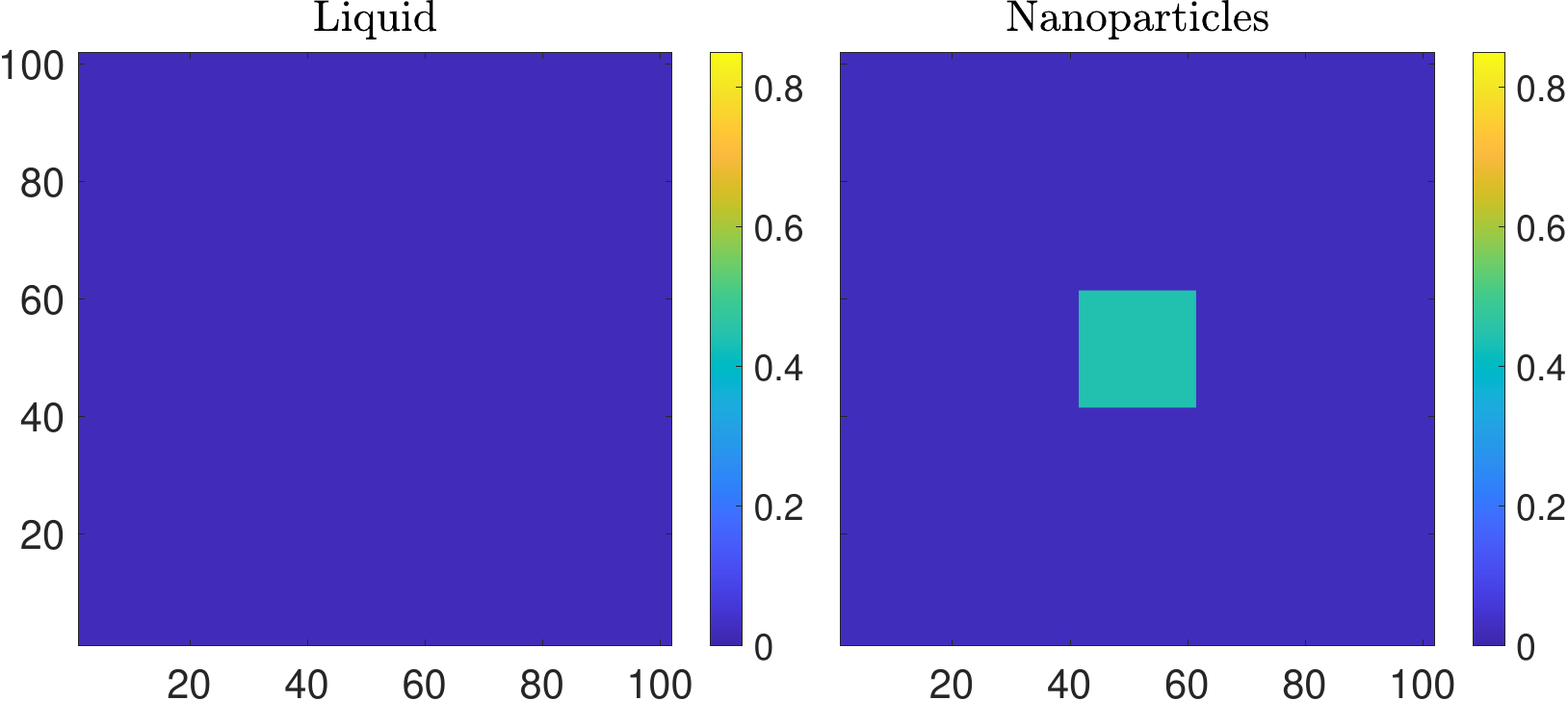}
\includegraphics[width=0.45\textwidth]{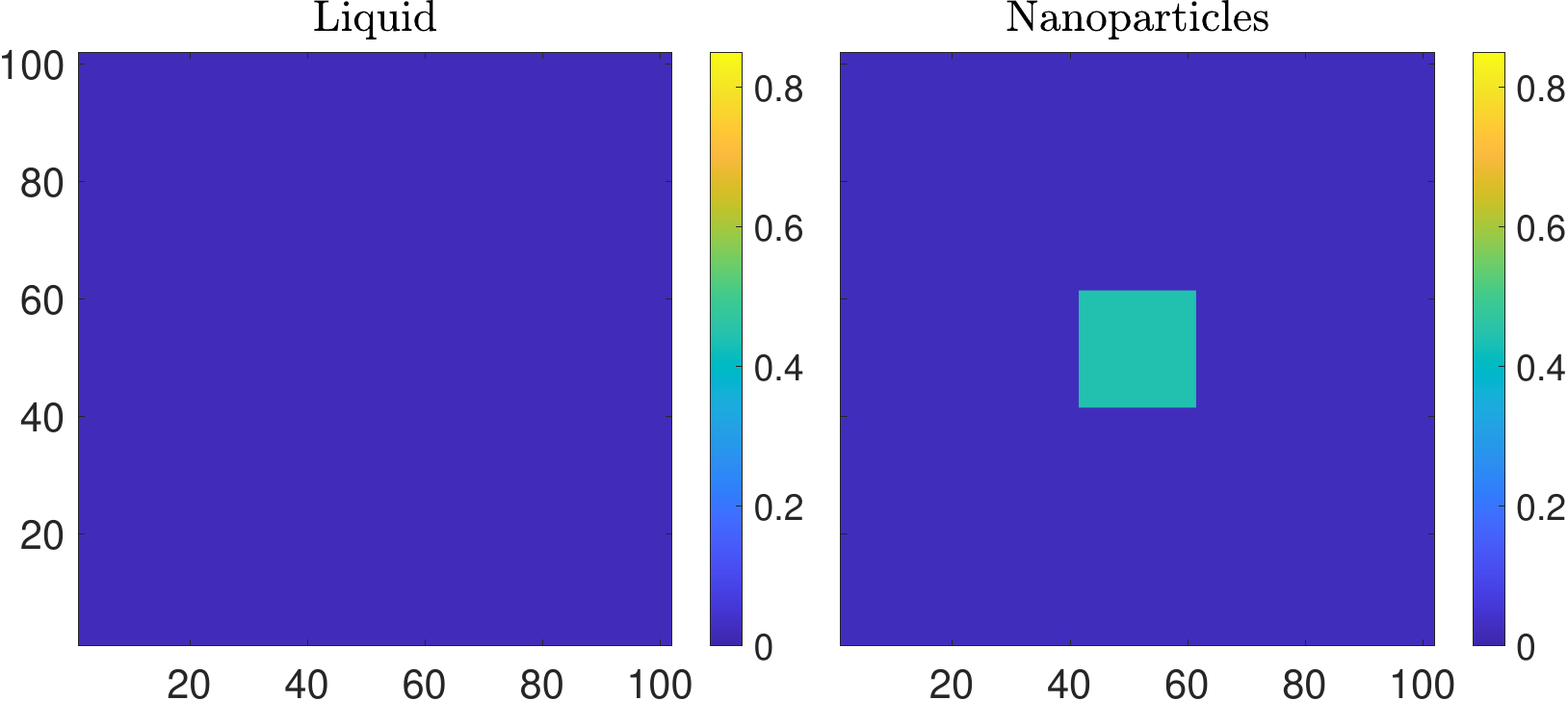}

\includegraphics[width=0.45\textwidth]{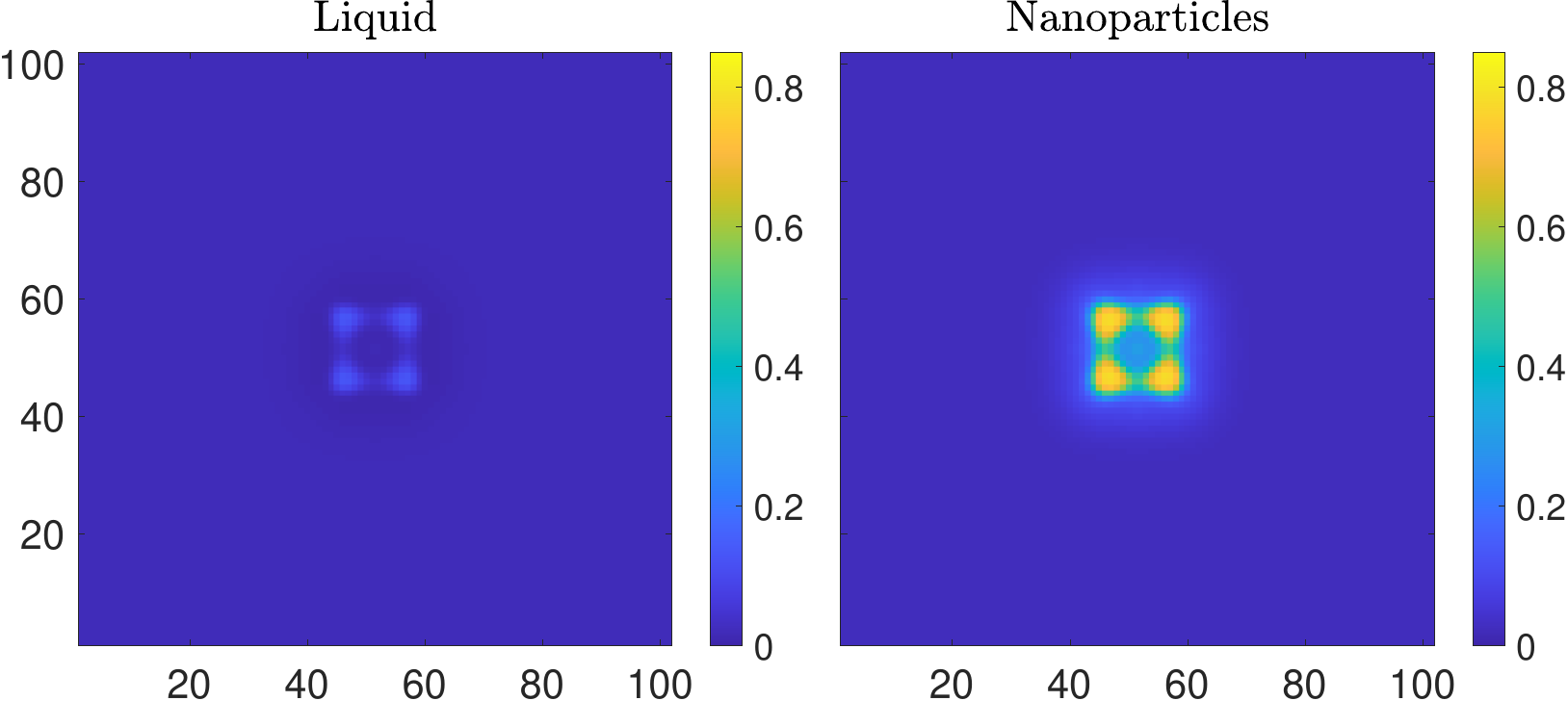}
\includegraphics[width=0.45\textwidth]{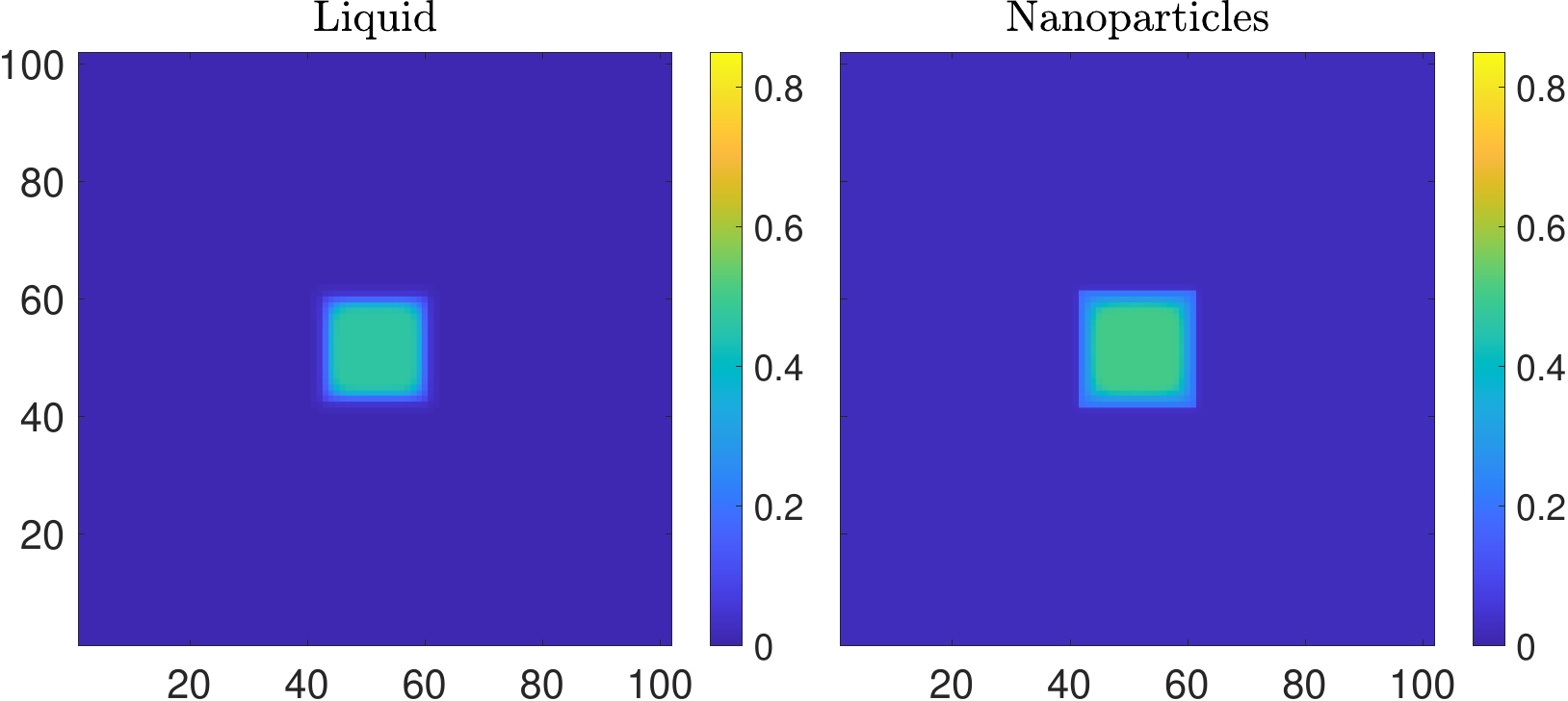}

\includegraphics[width=0.45\textwidth]{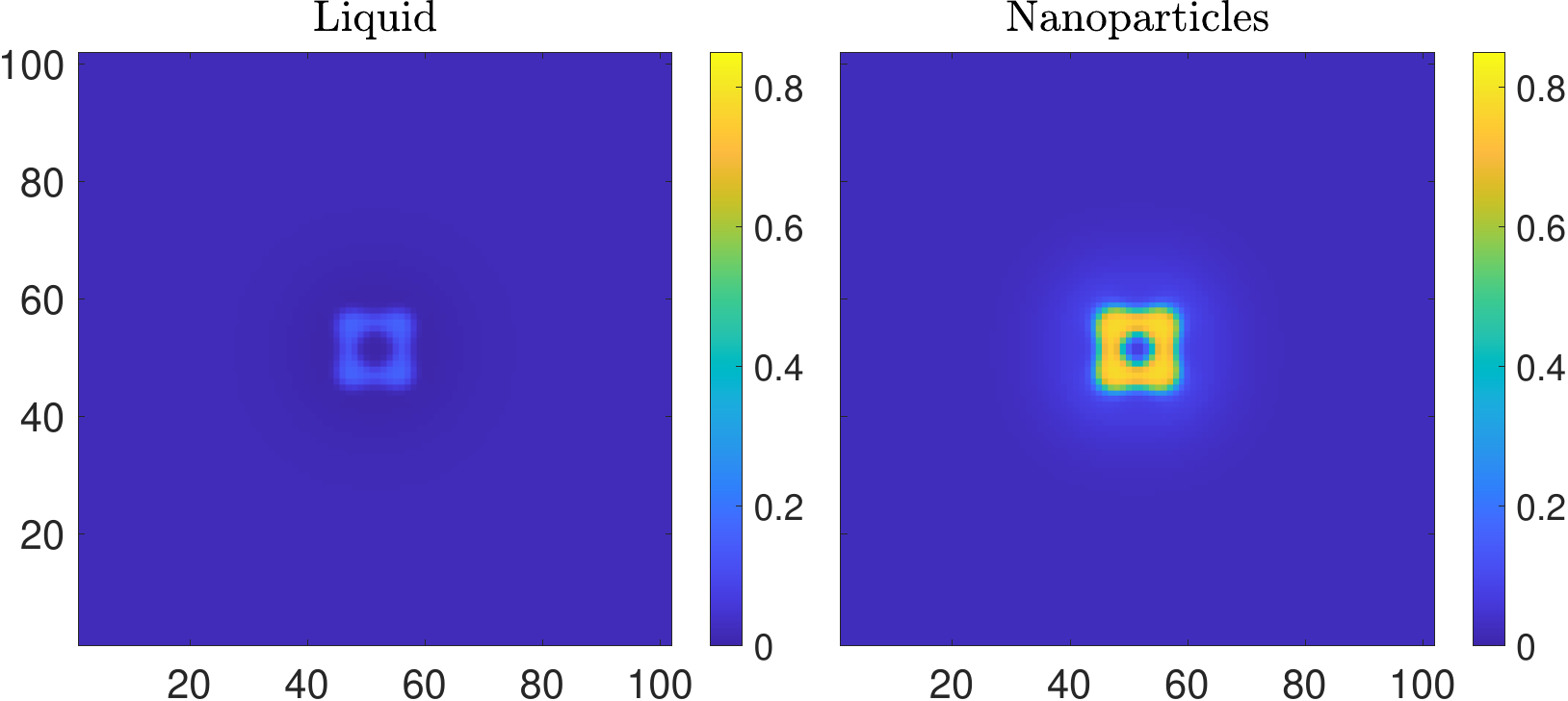}
\includegraphics[width=0.45\textwidth]{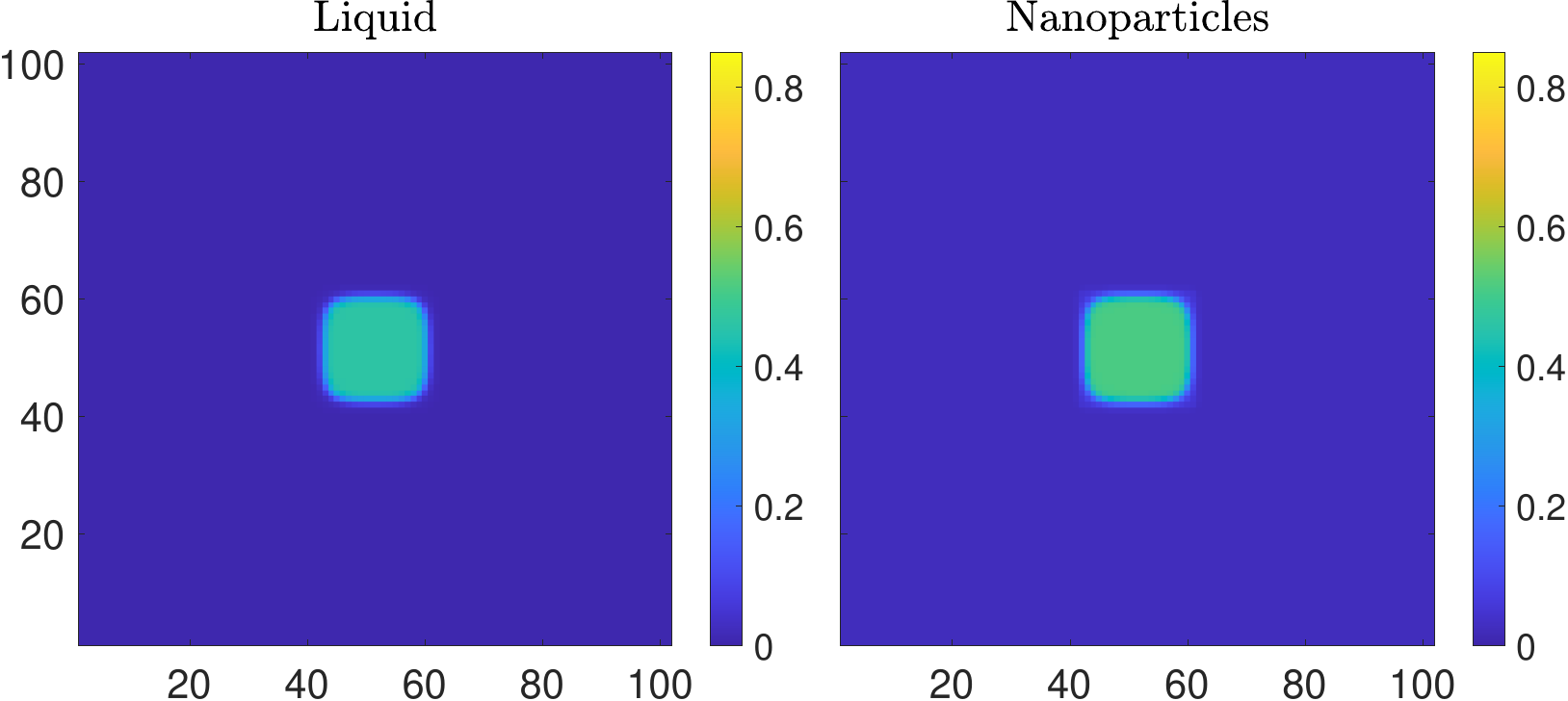}

\includegraphics[width=0.45\textwidth]{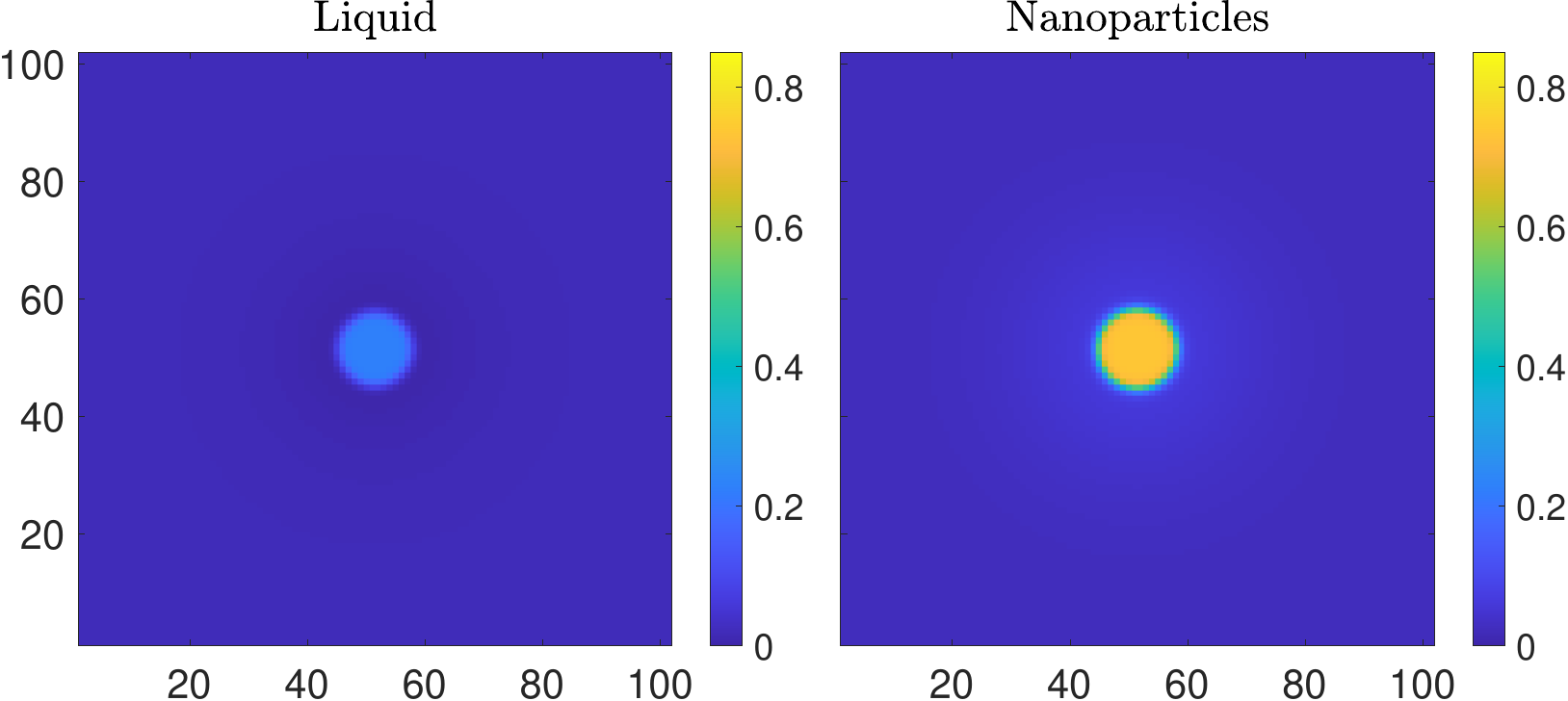}
\includegraphics[width=0.45\textwidth]{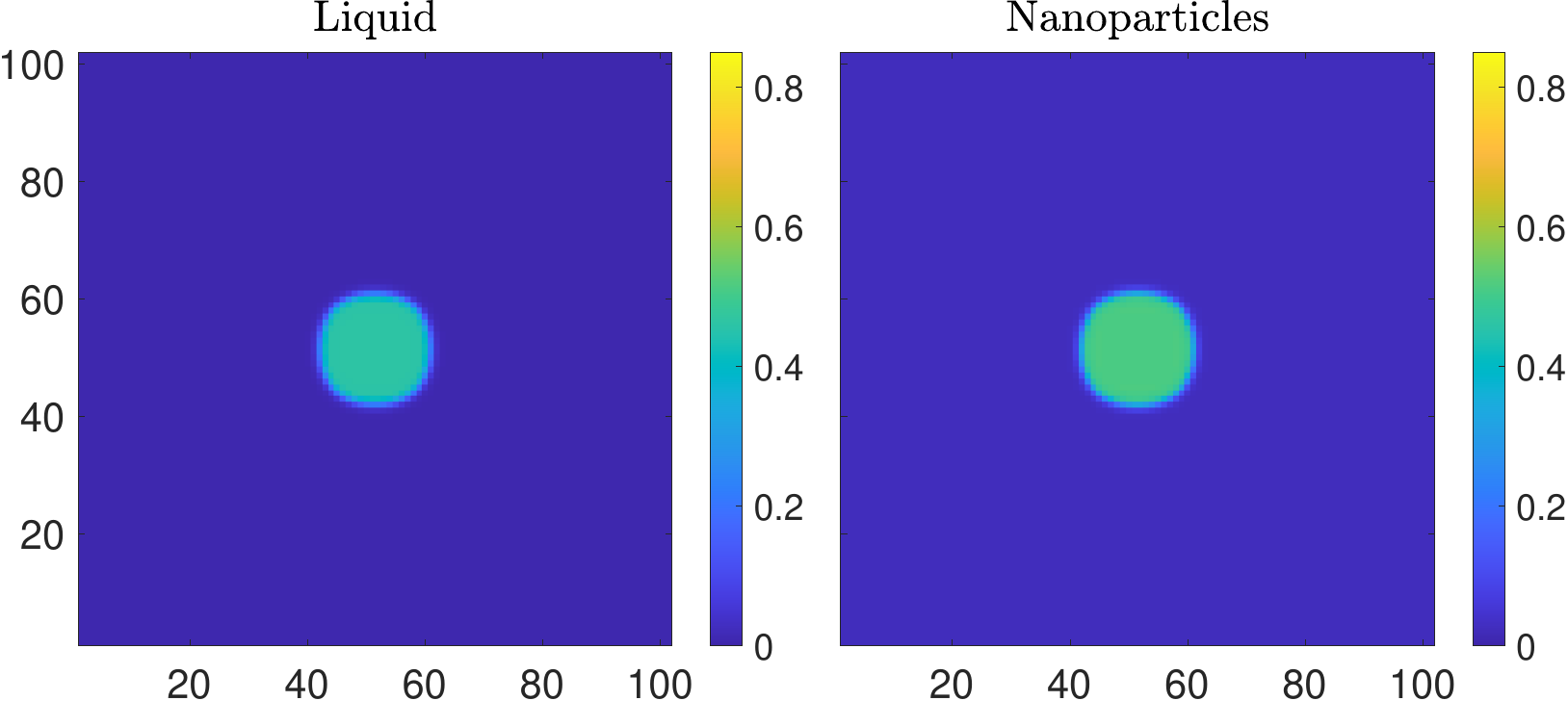}

\includegraphics[width=0.45\textwidth]{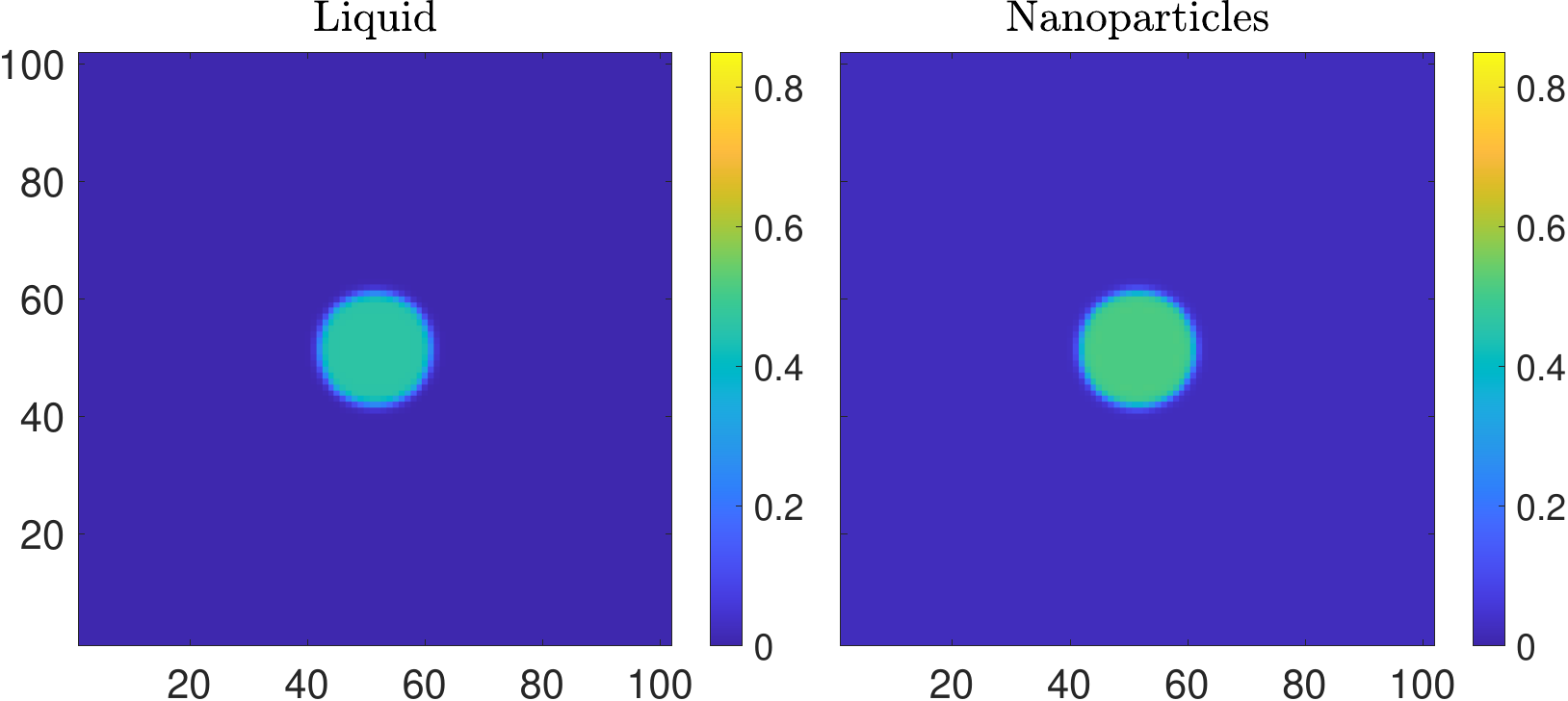}
\includegraphics[width=0.45\textwidth]{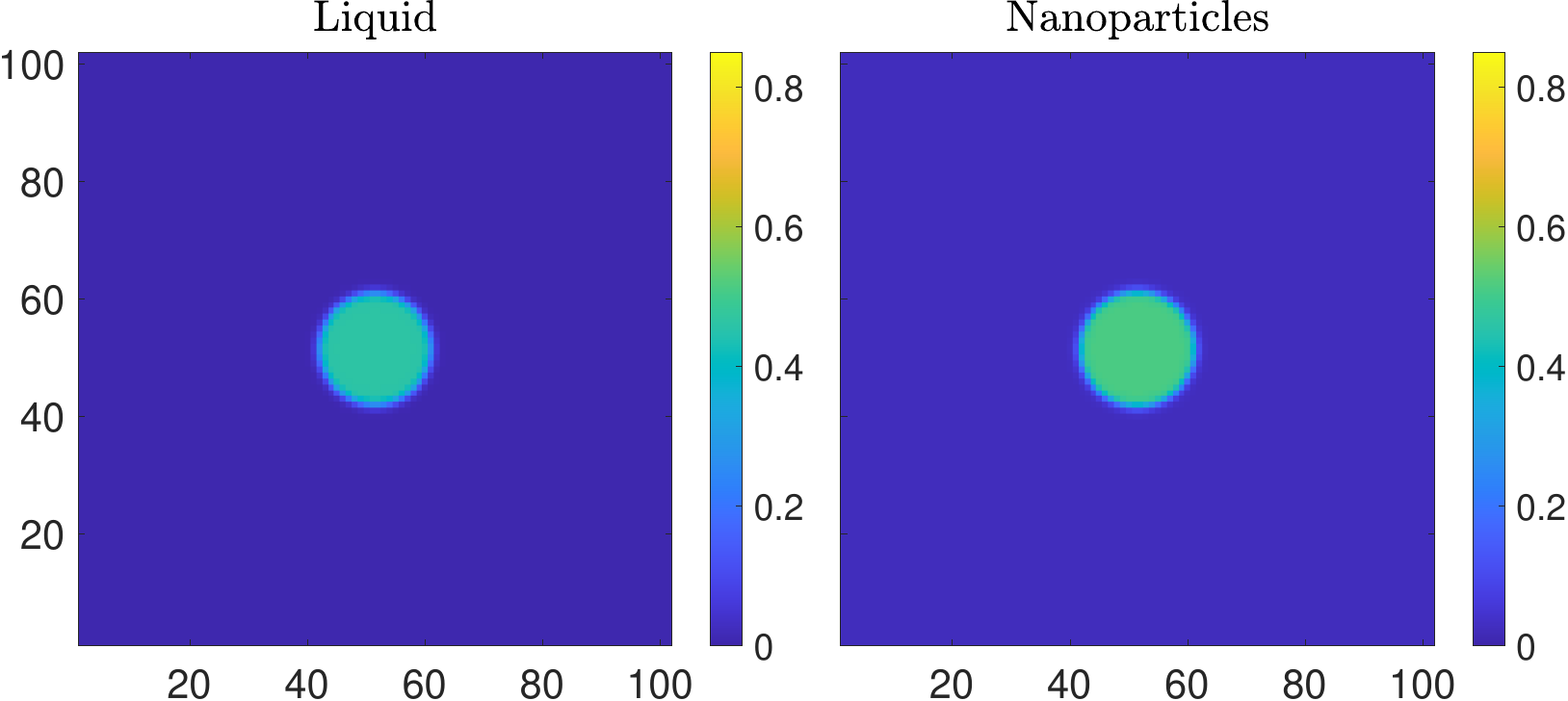}

\caption{Snapshots from a DDFT simulation (left) and from Picard iteration (a fictitious dynamics, right), starting from an initial condition where the liquid density is uniform and the nanoparticles are in a square region at the centre of the box. The initial average density of the liquid is selected so that the final states are the same as the equilibrium DFT calculation for $\beta\mu=-5$. The final equilibria from the two dynamics are the same, but the intermediate states during the evolution are very different. The DDFT profiles on the left are for the times $t=0$, 10, 30, 120 and $10^4$. The Picard profiles are from iterations 0, 200, 600, 2400 and 4000.
}
\label{fig:Picard_v_DDFT}
\end{figure}

In Fig.~\ref{fig:Picard_v_DDFT} we present results corresponding to an initial state where the density of the liquid is uniform throughout the system, while the the nanoparticles are gathered within a square region in the centre of the box.  We then perform both the DFT Picard minimisation and the DDFT simulation (with periodic boundary conditions for both the fluid and nanoparticles).  The initial and final average densities of the liquid in the box are the same for both systems. For the DDFT, the average density is a conserved quantity throughout the dynamics. For the Picard iteration, in this case it is roughly constant, but more generally this fictitious dynamics does not preserve mass between iterations. \red{Our Picard iteration results are obtained with mixing parameter $\alpha=0.01$ \cite{roth2010fundamental, hughes2014introduction}.} Interestingly, the results from these two approaches have very different paths to (the same) equilibrium.  This demonstrates that in this case the `quasi-dynamics' generated by the Picard scheme is not a good approximation of the DDFT dynamics. We see some rather striking transient states during the (realistic) DDFT dynamics displayed on the left of Fig.~\ref{fig:Picard_v_DDFT}, where the initial square block of nanoparticles breaks up into four smaller droplets that then subsequently re-coalesce into the single final droplet state. This complex dynamics is driven by a competition between bulk and interfacial contributions to the free energy, with each dominating at different stages of the dynamics.

\section{Concluding remarks}
\label{sec:conc}

We have presented a simple capillarity-approximation based theory for the size of nanoparticle laden liquid aerosol droplets. Our theory predicts how the size of the droplets varies depending on the vapour temperature, humidity, number of nanoparticles within the droplet and also the nature of the interactions between the nanoparticles and the liquid. We have validated our simple theory by comparing it with results from DFT. Our lattice DFT theory yields the density distribution of the particles within the aerosol droplets in addition to all the relevant thermodynamic quantities, such as the changes in the liquid-vapour interfacial tension due to varying concentrations of nanoparticles within the droplet. We have also developed a DDFT model, able to describe complex dynamical phenomena, such as droplet coalescence. \red{Concrete examples of the types of airborne aerosol systems that our model can be applied to include determining the stability (and therefore the lifetime) of exhaled droplets that can lead to the spread of COVID and other diseases, aerosol based therapies such as biomolecule inhalation therapy \cite{roudini2023acoustic}, crop spraying, and the myriad of different aerosols playing important roles in the world's atmosphere \cite{stevens2009untangling, wei2018aerosol, von2015chemistry}.}

Our simple CA model is useful for quick estimation of droplet sizes as a function of particle loading. For more precise calculations, our CA model could easily be improved by replacing the simple lattice-gas free energy \eqref{eq:f_bulk} used here with a more accurate equation of state. For example, the Mansoori-Carnahan-Starling-Leland equation of state for hard-sphere mixtures \cite{mansoori1971equilibrium, roth2010fundamental} could easily be used instead, or one of the many other accurate bulk fluid equations of state that are available in the literature -- see e.g.\ Refs.~\cite{chapman1989saft, kontogeorgis2006ten, gross2006equation}. The choice of a particular equation of state would be guided by obtaining additional information about the precise form of the molecular interactions in the system.

For those interested in a more accurate description of the density distribution of the liquid and nanoparticles within droplets, as future work one could replace the lattice DFT used here with a theory based on the lattice DFT of Refs.~\cite{maeritz2021density, maeritz2021droplet}, which gives an improved approach for dealing with the nearest-neighbour attractions between particles on a lattice. Alternatively, one could improve the model by using an accurate continuum DFT \cite{hansen2013theory}, such as a DFT based on fundamental measure theory \cite{rosenfeld1989,roth2010fundamental}. Such an approach would give a much better description of the liquid structure within droplets. This approach may be needed especially in cases where particles aggregate at the droplet liquid-vapour interface. For example, size-selectivity can occur in the drying of colloidal films containing two sizes of nanoparticles \cite{fortini2016dynamic, he2021dynamical, kundu2022dynamic}, so such effects may occur in the drying of aerosol droplets containing particle mixtures, potentially resulting in highly structured final states \cite{liu2019segregation}.

\section*{Acknowledgements}

This research was funded by the London Mathematical Society, the International Centre for Mathematical Sciences, and Loughborough University Institute of Advanced Studies. We are grateful to Emiliano Renzi and David Sibley for valuable discussions.

%\bibliography{refs}%{}
%\bibliographystyle{plain}


\end{document}